\documentclass[acmsmall]{acmart}
\AtBeginDocument{%
  \providecommand\BibTeX{{%
    \normalfont B\kern-0.5em{\scshape i\kern-0.25em b}\kern-0.8em\TeX}}}

\setcopyright{acmcopyright}
\copyrightyear{2018}
\acmYear{2018}
\acmDOI{XXXXXXX.XXXXXXX}

\acmJournal{JACM}
\acmVolume{37}
\acmNumber{4}
\acmArticle{111}
\acmMonth{8}
\usepackage{multirow}
\usepackage{subcaption}

\begin{document}

\title{A Review of Link Prediction Applications in Network Biology}

\author{Ahmad F. Al Musawi}
\affiliation{%
 \institution{Department of Computer Science, Virginia Commonwealth University}
 \city{Richmond}
 \state{Virginia}
 \country{USA}}
 \email{almusawiaf@vcu.edu}
\affiliation{%
  \institution{Department of Information Technology, University of Thi Qar}
  \city{Annasiriyah}
  \state{Thi Qar}
  \country{Iraq}}
\email{almusawiaf@utq.edu.iq}
\orcid{0000-0002-5508-9968}

\author{Satyaki Roy}
\affiliation{%
  \institution{Department of Mathematical Sciences, University of Alabama in Huntsville}
  \state{Alabama}
  \country{USA}
}
\email{sr0215@uah.edu}

\author{Preetam Ghosh}
\affiliation{%
 \institution{Department of Computer Science, Virginia Commonwealth University}
 \city{Richmond}
 \state{Virginia}
 \country{USA}}
 \email{pghosh@vcu.edu}







\renewcommand{\shortauthors}{Al Musawi, Roy, and Ghosh}

\begin{abstract}

In the domain of network biology, the interactions among heterogeneous genomic and molecular entities are represented through networks. \textit{Link prediction} (LP) methodologies are instrumental in inferring missing or prospective associations within these biological networks. In this review, we systematically dissect the attributes of local, centrality, and embedding-based LP approaches, applied to static and dynamic biological networks. We undertake an examination of the current applications of LP metrics for predicting links between diseases, genes, proteins, RNA, microbiomes, drugs, and neurons. We carry out comprehensive performance evaluations on established biological network datasets to show the practical applications of standard LP models. Moreover, we compare the similarity in prediction trends among the models and the specific network attributes that contribute to effective link prediction, before underscoring the role of LP in addressing the formidable challenges prevalent in biological systems, ranging from noise, bias, and data sparseness to interpretability. We conclude the review with an exploration of the essential characteristics expected from future LP models, poised to advance our comprehension of the intricate interactions governing biological systems.

\end{abstract}



\received{20 February 2007}
\received[revised]{12 March 2009}
\received[accepted]{5 June 2009}

\maketitle

\section{Introduction}

Understanding the intricate interplay among biological entities lies at the core of many questions in the field of systems biology. One approach to addressing this challenge involves harnessing the power of network analysis, notably through the mathematical construct of \textit{graphs}, where biological entities are represented as nodes, and their interconnections take the form of links~\cite{pavlopoulos2011using,berger2008graph}. The integration of network theory in the study of molecular and biomedical interactions has led to the emergence of \textit{network biology}. This interdisciplinary domain harnesses cutting-edge computational and visualization methodologies to unravel the intricate structures and dynamic behaviors intrinsic to complex biological systems based on the underlying molecular interactions, namely, protein-protein, metabolic, signaling, and transcription-regulatory interactions~\cite{barabasi2004network,BARH2014385}.

The concept of \textit{link prediction} (LP) plays a pivotal role in the burgeoning field of network biology. It involves forecasting the likelihood of missing or unknown connections or interactions between entities in a network~\cite{lu2011link}. LP has been applied successfully to build recommender systems for social networks, collaborative filtering, viral marketing, etc, based on similarity in individual profiles~\cite{su2020link}. In the realm of network biology, this translates to the capability to anticipate relationships among biological entities, such as genes, proteins, or metabolites. These predictive techniques rely on the analysis of existing network topology, incorporating knowledge from diverse sources like protein-protein interaction databases, gene expression profiles, or disease association data~\cite{breit2020openbiolink}. 

The applications of link prediction in network biology are manifold. They enable the identification of potential protein-protein interactions, the inference of gene regulatory networks, the prediction of disease-gene associations, and the exploration of underlying biological pathways \cite{mirsig,rev1,rev2}. By unveiling these hidden connections, link prediction contributes significantly to our understanding of complex biological systems, facilitating the discovery of novel biomarkers, drug targets, and insights into the intricate web of biological interactions. Furthermore, LP techniques can help researchers model and predict these temporal dynamics by forecasting how network connections will develop or weaken over time~\cite{divakaran2020temporal}. This temporal understanding is particularly crucial for tracking disease progression, understanding gene expression changes, or following the evolution of biological pathways.

\begin{figure}[h!]
    \centering
    \includegraphics[width=3.25in]{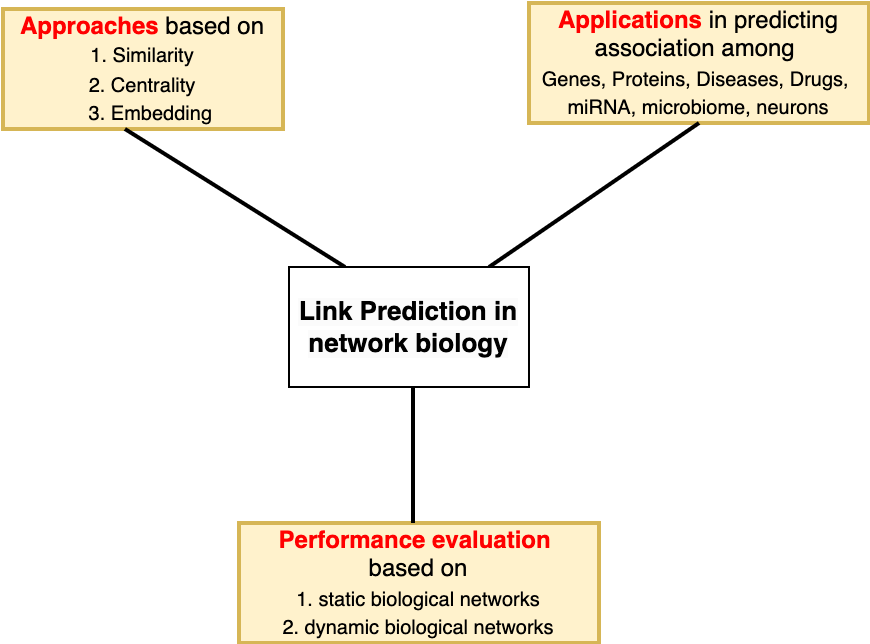}
    \caption{Outline of the review on link prediction applications in network biology}
    \label{fig:outline}
\end{figure}

In this review, we delve into the application of link prediction (LP) techniques within the domain of biological networks (refer to Fig. \ref{fig:outline}). Our exploration spans various classes of LP methods, encompassing those reliant on immediate neighborhoods, network centrality, and embedding techniques. We focus on their effectiveness in predicting associations among various biological entities, including diseases, genes, proteins, and drugs, to name a few. Our investigation extends to the evaluation of performance metrics on both static and dynamic biological networks. Finally, we synthesize our findings with a discussion of the challenges encountered and future avenues of research in the ongoing journey of LP's application in the intricate landscape of network biology.

This paper has been organized as follows. Sec. 2 covers the preliminaries of graphs and their features leveraged to study biological systems. Sec. 3 covers the three classes of LP techniques, namely, similarity-based, centrality-based, and embedding-based, while Sec. 4 presents its existing applications in network biology. Sec. 5 demonstrates the performance of LP in static and dynamic biological networks. Finally, Sec. 6 and 7 discuss the challenges, future directions, and conclusions.

\section{Preliminaries}
A biological network is represented as a graph $G (V, E)$, where the vertex (or node) set $V$ may represent genes, transcriptional factors, proteins, diseases, drugs or side effects, etc. and the links represent an association between a pair of nodes. A graph $G$ can be represented as an adjacency matrix $A$ such that $A_{uv} = 1$ if there is a relationship between nodes $v_x, v_y$ and $A_{uv} = 0$ otherwise. 

\subsection{Types of Graphs}\label{sec:type}

Different types of graphs are used to capture myriad biological interactions~\cite{west2001introduction}. The edges of $G$ may be \textit{directed} or \textit{undirected}. In the undirected graph, the relationship between two nodes is symmetric, i.e., $A_{ij}=A_{ji}$, while in a directed graph, edges have a direction (asymmetric) where $A_{ij} \neq A_{ji}$. The edges can also be \textit{weighted} or \textit{unweighted}. In a weighted graph, the edge weights measure the strength or intensity of the relationship between nodes while all edges have the same weight in an unweighted graph. On the other hand, links may have \textit{signs}, namely, positive and negative. where positive and negative edges may denote the up-regulation and down-regulation of a gene by a transcription factor, respectively. Finally, a \textit{homogeneous} graph, also called a non-attributed graph, consists of nodes of a single type, whereas \textit{heterogeneous} graphs contain nodes and edges of different types, allowing for greater information expression. Heterogeneous graphs can be further categorized into bipartite, tripartite, or multimodal, depending on the number of node types.

A \textit{bipartite} graph is effective when modeling the relationship between a pair of biomedical entity types, say disease set ($a \in A$) and drug set ($b \in B$). A bipartite network contains disjoint sets of nodes $A$ and $B$, where a possible link $(a, b)$ may only exist between a disease and a drug. On many occasions, a bipartite network is converted from being a \textit{two-mode} network into a \textit{one-mode} projection network comprising a single entity set as nodes. In other words, the drug-disease bipartite network is converted into a network containing diseases as nodes and links between a pair of diseases sharing one or more drugs. A bipartite graph can be generalized to $k$-partite graphs to represent $k$ biomedical entity types. Lastly, graphs can evolve over a period of time resulting in temporal networks. Let $\mathbf{G} = \{G_1,..., $G$_T\}$ be a set of networks that represent the evolving behavior of a graph over different $T$ time steps. Evolution typically refers to the emergence or disappearance of edges among the set of nodes, keeping the vertex set unaltered, i.e., $\mathbf{G} = \{(V, E_t)| t \in \{1,.., T\}\}$. 


\subsection{Problem setting}\label{sec:setting}
An input network dataset represented by a graph $G = (V, E)$ is divided into two subgraphs of training $G^{\tau} = (V, E^{\tau})$ and testing $G^P (V, E^{P})$, which have non-overlapping edge sets. Training graph $G^{\tau}$ is created by randomly sampling a fraction $f$ of all links, i.e., $|E^{\tau}| = f \times |E|$, The testing (or probe) graph $G^P$ is obtained by taking the remaining edges $E^P$ from $G$, making $E = E^{\tau} \cup E^P$. The link prediction (LP) model is trained on $G^{\tau}$ before predicting the likelihood of an edge between node pairs in $G^P$. Finally, the accuracy of the LP model is high if it assigns a high likelihood score to the existing links in $G^P$ and low scores to nonexistent links, and vice versa.

\section{Methodology}\label{sec:method}
We provide an overview of the link prediction methods that have been widely used for predicting links in the biomedical and biological domains. These methods are categorized into local similarity and path-based approaches, centrality-based approaches, and representation learning approaches. 

\subsection{Similarity-based Approaches}~\label{sec:sim}

One intuitive and widely used approach for predicting potential relationships is to look over the similarity of features shared between any unconnected pair of nodes. 

\subsubsection{Local approaches}\label{sec:local}
The local approaches explore the immediate neighborhood of a node $u$, comprising a set of nodes directly connected to them within a network (denoted by $\Gamma_u$). The property that \textit{two nodes have greater potential to share a relationship (i.e., a link) if they share similar neighbors} has been leveraged by several approaches. The common neighbor (CN) metric gauges the likelihood of a link between nodes $u, v$ as the number of neighbors they share~\cite{lorrain1971structural}, i.e., $S^{CN}_{u,v} = |\Gamma_u \cap \Gamma_v|$. The variants of this approach are summarized in Table \ref{tab:local} and have been used as baselines for the prediction of biological relationships (refer to Sec. \ref{sec:apply} for details).

\begin{table}[h!]
\caption{Unweighted and weighted local similarity-based metrics:  $\bar k_u$ is the weighted degree of node $u$, $\lambda_u = max(0, n_{avg} - \bar k_u)$, $n_{avg}$ represents weighted average node degree of the network, wLCL is the sum of the weights of the links connecting the common neighbors of $u$ and $v$ (local community links). $e_u$ refers to the external degree of $u$, computed considering the neighbours of $u$ that are not common neighbours of $u$ and $v$.}

\centering
\begin{tabular}{p{7cm} p{6cm}} 
 \hline
 \hline 
 \textbf{Unweighted metric} & \textbf{Formula} \\ 
 Salton (cosine similarity, CS)~\cite{salton1983introduction} & $S^{SA}_{u,v} = \frac{|\Gamma_u \cap \Gamma_v|}{\sqrt{|\Gamma_u||\Gamma_v|}}$ \\ 
 Jaccard Index (JI)~\cite{jaccard1901etude} & $S^{JA}_{u,v}~ = \frac{|\Gamma_u \cap \Gamma_v|}{|\Gamma_u \cup \Gamma_v|}$\\
 Sorenson index (SI)~\cite{sorensen1948method} & $S^{HPI}_{u,v} = \frac{|\Gamma_u \cap \Gamma_v|}{min(|\Gamma_u|, |\Gamma_v|)}$\\
 Hub Promoted Index (HPI)~\cite{ravasz2002hierarchical} & $S^{JA}_{u,v}~ = \frac{|\Gamma_u \cap \Gamma_v|}{|\Gamma_u \cup \Gamma_v|}$\\
 Hub Depressed Index (HDI)~\cite{zhou2009predicting} & $S^{HDI}_{u,v} = \frac{|\Gamma_u \cap \Gamma_v|}{max(|\Gamma_u|, |\Gamma_v|)}$\\
 Resource Allocation (RA)~\cite{zhou2009predicting}& $S^{RA}_{u,v} = \sum\limits_{z \in \Gamma_u \cap \Gamma_v} \frac{1}{|\Gamma_z|}$\\
 Leicht-Holme-Newman(LHN-1)~\cite{leicht2006vertex} & $S^{LHN}_{u,v} = \frac{|\Gamma_u \cap \Gamma_v|}{|\Gamma_u| |\Gamma_v|}$\\
 Preferential Attachment (PA)~\cite{barabasi1999emergence} & $S^{PA}_{u,v} = |\Gamma_u| |\Gamma_v|$\\
 
 Adamic-Adar (AA)~\cite{adamic2003friends} & $S^{AA}_{u,v} = \sum\limits_{z \in \Gamma_u \cap \Gamma_v} \frac{1}{\log |\Gamma_z|}$\\
 
 Cannistraci-Alanis-Ravasi-based variation of the resource allocation (CAR)~\cite{cannistraci2013link,ghasemian2020stacking} & $S^{CAR}_{u,v} = \sum\limits_{z \in \Gamma_u \cap \Gamma_v} \frac{\Gamma_u \cap \Gamma_v \cap \Gamma_z}{|\Gamma_z|}$\\
 
 Cannistraci-Jaccard (CJC) \cite{daminelli2015common}&
 $S^{CJC}_{u,v} = \frac{|\Gamma_u \cap \Gamma_v|. S^{LCL}_{u,v}}{\Gamma_u \cup \Gamma_v}$\\
 
 Cannistraci–Adamic–Adar (CAA) \cite{daminelli2015common}&
 $S^{CAA}_{u,v} = \sum\limits_{z \in (\Gamma_u \cap \Gamma (\Gamma_v)) \cup  (\Gamma_v \cap \Gamma (\Gamma_u)) } \frac{|\gamma(z)|}{\log |\Gamma_z|}$\\

CPA \cite{daminelli2015common, cannistraci2013link}&
 $S^{CPA}_{u,v} = e_u e_v + e_u S^{CAR}_{u,v} + e_v S^{CAR}_{u,v} +  (S^{CAR}_{u,v})^2$\\

 CRA \cite{daminelli2015common}&
 $S^{CRA}_{u,v} = \sum\limits_{z \in (\Gamma_u \cap \Gamma (\Gamma_v)) \cup  (\Gamma_v \cap \Gamma (\Gamma_u)) } \frac{|\gamma(z)|}{|\Gamma_z|}$\\
 
 Mutual Information Index (MI) \cite{tan2014link}&
 $
 S^{MI}_{u,v} = \sum\limits_{k \in O_{uv}} I(L^1_{uv};k)-I(L^1_{uv})
 $
 \\
\hline
 \textbf{Weighted metric}  & \textbf{Formula} \\ 
 Common Neighbors (wCN) & $ wCN_{u,v}= \sum_{z \in \Gamma_u \cap \Gamma_v} \frac{w(x,z)+w(y,z)}{2}$\\

Preferential Attachment (wPA) & $wPA_{u,v} = \bar k_u \times \bar k_y$\\

 Adamic-Adar (wAA) & $wAA_{u,v} = wCN_{u,v} \times \frac{1}{log(\sum_{\bar z \in \Gamma(z)} w(\bar z, z))}$\\

 Adjusted Czekanowski-Dice Dissimilarity (wACDD) & $wACDD_{u,v} = \frac{2 wCN_{u,v}}{\bar k_u + \bar k_y + \lambda_u + \lambda_v}$\\

 Cannistraci-Alanis-Ravasi Index (wCAR) & $wCAR_{u,v} = wCN_{u,v} \times wLCL$\\
 \hline
 \hline

\end{tabular}
\label{tab:local}
\end{table}

\subsubsection{Path-based approaches}\label{sec:path}
These measures explore a slightly larger neighborhood (as opposed to the one-hop neighborhood as is the case with local approaches) and similarity is measured by the degree of overlap in the larger neighborhoods~\cite{muscoloni2018local, zhou2021experimental}. 

\begin{enumerate}
    \item \textit{CH2-L2 Index} is a link prediction model that assigns a reward for the internal connectivity existing among common neighbors and penalizes outside connectivity. 
    \begin{equation}
        S^{CH2-L2}_{u,v} = \sum \limits_{i \in \Gamma_u \cap \Gamma_v} \frac{1 + C_i}{1 + O_i}
    \end{equation}

    Here $C_i$ represents the number of neighbors of node $i$ that exist in  $\Gamma_u \cap \Gamma_v$, $O_i$ represents the number of neighbors of node $i$ that do not exist in $ \Gamma_u \cap \Gamma_v$ nor in $u$ or $v$.
    
    \item \textit{CH2-L3 Index}: very similar to CH2-L2 metric, this metric considers all three path lengths (two intermediate nodes $u,v$) between the targeted edge ($u, v$). 
    
    \begin{equation}
        S^{CH2-L3}_{u,v} = \sum \limits_{i \in \Gamma_u, j \in  \Gamma_v} \frac{A_{i,j} \sqrt{(1 + \Bar{C_i})(1 + \bar{C_j})}}{\sqrt{(1 + \bar{O_i})(1 + \bar{O_j})}}
    \end{equation}
    
    Here $\bar{C_i}$ represents the number of links between node $i$ and all the nodes that exist in the set of intermediate nodes on all 3-hop paths connecting nodes $u$ and $v$, $\bar{O_i}$ represents the number of links between node $i$ and all nodes that are not $u, v$ nor the intermediate nodes on any 3-hop paths connecting $u$ and $v$.
\end{enumerate}

\begin{table}[h!]
\caption{Network centrality measures, where $m$ is the sum of the weights of all edges, $A$ is an adjacency matrix, $k_i$ is the degree of node $i$, $c_i$ is the community of node $i$, $\delta$ is 1 if $c_i = c_j$ and 0 if they are not. $d(v,u)$ is the shortest-path distance between node $u$ and node $v$. $\sigma_{st}$ is the number of shortest paths from $s$ to $t$, $\sigma_{st}(v)$ is the number of shortest paths from $s$ to $t$ that pass through $v$. $V_{NB\_INT}(u)$ is the neighbor node set of source node $u$, $E_{NB\_INT}(u)$ is the edge set in $G_{NB}(u) = (V_{NB}(u),E_{NB}(u))$.  $\alpha$ and $\beta$ are constants and $d_k$ is the out-degree of node $k$ if such degree is positive, or $d_k = 1$ if the out-degree of $k$ is null.}

\centering
\begin{tabular}{p{3cm} p{4.5cm} p{5.5cm}} 
 \hline
 \textbf{Metric} & \textbf{Description} & \textbf{Formula} \\ 

  Modularity~\cite{newman2006modularity} & Find network communities &
$Q = \frac{1}{2m} \sum\limits_{uv \in V} [A_{uv} - \frac{k_u k_v}{2m}] \delta(c_u.c_v)$ \\

Closeness \cite{bavelas1950communication} & How close a node is to all other nodes& $C(u) = \frac{N - 1}{\sum_{v \neq u} d(v, u)}$\\

 Betweenness~\cite{brandes2001faster} & Nodes that intercept paths between other nodes &
 ${BC}_v = \sum\limits_{s\neq v \neq t \in V} \frac{\sigma_{st}(v)}{\sigma_{st}}$\\

 Katz~\cite{katz1953new}& Nodes that reach other community nodes & $x_u = \alpha \sum\limits_v A_{uv} x_v + \beta$\\

 Eigenvector and Pagerank~\cite{bonacich1987power,page1999pagerank} & Nodes connected to other important nodes & $Ax = \lambda x$; $x_u = \alpha \sum\limits_k \frac{A_{k,u}}{d_k} \, x_k + \beta$\\

 Hubs and authority~\cite{kleinberg1999authoritative}& Nodes that are well-connected (hubs) and nodes that bridge hubs (authority) & $auth(u) = A^T hub(u)$ $hub(u) = A auth(u)$\\

 Local interaction density~\cite{luo2015identification} & Nodes with high interaction among its neighbors & $LID(u) = \frac{|E_{NB\_INT}(u)|}{|V_{NB\_INT}(u)|}$ \\
 
 Linear Threshold Rank~\cite{riquelme2019neighborhood} & Nodes with ability to influence non-neighbors & $LTR(i) = \frac{|f(\{i\}) \cup \Gamma(i)|}{n}$\\
  

 \hline
\end{tabular}

\label{tab:centrality}
\end{table}

\subsection{Centrality-based Approaches}\label{sec:central}

Network centrality measures (refer to Table \ref{tab:centrality} for details) quantify the importance of nodes within a network. These measures are based on the following concepts in network theory. 

\begin{itemize}

    \item \textit{Community influence}: Network modularity~\cite{newman2006modularity} is a standard metric to identify network modules or communities. The relative importance of any node within a community is measured using Katz centrality capturing the reachability of other community nodes to the given node via its immediate neighbors or first-degree nodes~\cite{katz1953new}. Local interaction density (LID) estimates the amount of interaction between the neighbors of a given node~\cite{luo2015identification}. Similarly, the edge clustering coefficient (ECC)~\cite{mistry2017diffslc} for an edge connecting nodes \( u \) and \( v \) is:

    \begin{equation}
        ECC^{(3)}_{u,v} = \frac{z^{(3)}_{u,v} + 1}{\min\{(k_u - 1), (k_v - 1)\}}
    \end{equation}
    
    Here, \( z^{(3)}_{u,v} \) denotes the total number of triangles that encompass the edge \( (u,v) \), while \( k_u \) and \( k_v \) represent the number of triangles that the nodes \( u \) and \( v \) each participate in, respectively.
    
    \item \textit{Neighbors of significance}: The importance of a node is measured in terms of the importance of its first-degree nodes. Both PageRank and eigenvector centralities deem a node significant if other important nodes point to it~\cite{page1999pagerank,bonacich1987power}. For example, the common neighbor and centrality-based parameterized algorithm (CCPA)~\cite{ahmad2020missing} merges common neighbors with node centrality, particularly, closeness centrality, which is defined by the average shortest path to other nodes.

    \begin{equation}
        S_{u,v}^{CCPA} = \alpha . (|\Gamma_u \cap \Gamma_v|) + (1-\alpha).\frac{|V|}{d_{uv}}
    \end{equation}
    
    The term $\frac{|V|}{d_{uv}}$ represents the closeness centrality for edge $(u,v)$,  $d_{uv}$ is the shortest distance between the nodes $u$ and $v$. $\alpha$ represents the user-set parameter which falls between 0 and 1. It balances the influence of common neighbors and centrality.
    
    \item \textit{Bridging components}: A central node is one that acts as a bridge connecting different network components. Betweenness centrality measures the number of shortest paths between other node pairs that a given node intercepts~\cite{brandes2001faster}. Hub and authority centrality generalizes the Eigenvector centrality by finding well-connected nodes (termed hubs) and bridges between the hubs (termed authority)~\cite{kleinberg1999authoritative}.

    \item \textit{Influence diffusion}: Linear Threshold Rank (LTR) is defined as the total number of nodes activated by influence diffusion when the initial activation set is formed by a node and its immediate neighbors~\cite{riquelme2019neighborhood}.
    
\end{itemize}

\subsection{Representation Learning-based Approaches}\label{sec:embed}
In this section, we cover approaches that encode the structural properties of nodes into low-dimensional vector representations, called \textit{embeddings}, before predicting the existence of links between a pair of nodes based on the similarity in their embeddings. We classify these methods into three groups: matrix factorization-based, random walk-based, and neural network-based.

\subsubsection{Matrix Factorization based methods}\label{sec:MF}
Non-negative matrix factorization (NMF) \cite{wang2012nonnegative,cai2010graph} is a dimension reduction technique that decomposes a large data matrix $X = [x_1,...,x_N] \in \mathbb{R}^{M \times N}$ into a product of two (or more) smaller matrices, i.e., $X \approx UV^T$, $U = [u_{ik}] \in \mathbb{R}^{M \times K}, V = [v_{jk}] \in \mathbb{R}^{N \times K}$, where the number of latent features is a user-defined input. The decomposed matrices $U, V$, both of which are constrained to have non-negative entries, are termed \textit{feature matrix} and \textit{coefficient matrix}, respectively. While the feature matrix contains the embedding of each data point (or node in the context of graph data) to be leveraged for subsequent link prediction tasks, the coefficient matrix informs the contribution of a feature in the feature vector toward the data point. 

Specifically for graph data, the matrix $X$ could be a weighted or unweighted adjacency matrix. NMF starts by initializing $U$ and $V^T$ with random non-negative values. Since the goal of the optimization is to minimize the difference between $X$ and $UV^T$, several cost functions are proposed such as Frobenius norm (see Eq. \ref{Forbenius}), Kullback-Leibler (KL) divergence, and Itakura-Saito divergence. Also, several \textit{iterative algorithms} exist for updating both $U$ and $V$ to minimize the cost function, such as multiplicative updates, alternating least squares, and projected gradient descent. 

\begin{equation}
    \min_{U, V} ||X-UV^T||^2 = \min(\sum_{i,j} (x_{ij} - \sum_{k=1}^K u_{ik} v_{jk}^T)^2)
    \label{Forbenius}
\end{equation}

NMF lends itself to temporal link prediction approaches (see Sec. \ref{sec:temporal} for details), where the goal is to analyze the evolving interaction between a given pair of nodes over time~\cite{zhang2021temporal}. The following three variants have been put forth to meet different optimization requirements of NMF:

\begin{enumerate}

    \item In the simplest form, it minimizes the approximation error between the adjacency matrix $A$ and decomposed matrices $U, V$ in three consecutive time points, i.e., $\min \sum || A - U.V^T ||^2$. 

    \item The intrinsic topological properties of the network have been incorporated by refining the objective function as $\min \sum a_{u, v} \times || f_u - f_v ||^2$. This ensures that the latent vectors of two nodes $u$ and $v$ are similar if they are located nearby in the input network $G$.

    \item A consensus representation of the feature matrix, given by $U^{*}$ was incorporated to tackle the problem of outliers in learning the latent factor. Instead of learning the shared basis matrix of all snapshots $U$, $U^{*}$ is estimated from all previous snapshots. The direct application of these methods has been highlighted in the application section (see Sec. \ref{sec:app_dis_gene}).
        
\end{enumerate}


NMF has a wide range of applications in data mining, medical imaging, voice processing, and link prediction~\cite{chen2017link}. The other matrix factorization-based techniques are based on singular value decomposition~\cite{franceschini2016svd}, principal component analysis~\cite{bro2014principal}, and independent component analysis~\cite{du2013neural}. 

\subsubsection{Random walk-based methods}\label{sec:random}
Several graph-based learning models utilize the NLP deep learning concept of SkipGram in their models. SkipGram \cite{mikolov2013efficient} is a neural network model with one hidden layer aimed at learning the association of words given a large document collection. The inputs of the SkipGram model are the co-occurrence of words in the same sentence (window). This model preserves the topological proximity of the words into embedded representation.

The DeepWalk \cite{perozzi2014deepwalk} algorithm (DW) is a graph representation learning technique that involves generating random walks from each node in a graph and using the word2vec algorithm to learn the embeddings of the nodes. During the random walk generation step, a fixed-length walk is performed from each node, and the resulting sequence of nodes is treated as a sentence. The Skip-gram variant of the word2vec algorithm is then applied to learn the node embeddings, which capture the graph's structural relationships and provide a low-dimensional representation of the nodes. To determine the embeddings, DeepWalk introduces a mapping function $\phi: v \in V \rightarrow \mathbf{R}^\mathbf{d}$. Specifically, it leverages local information obtained from truncated random walks. The similarity between any two nodes $u,v$, given by $sim (u, v)$, is commensurate with the frequency of co-occurrence in the same walks. Finally, the embedding of node $u$ ($\phi_u$) preserves the topological similarity, as:

    \begin{equation}
        sim (u, v) = \phi_{u}^{T} \times \phi_v
    \end{equation}

\noindent DeepWalk generates random walks starting on each node in the graph to be used to gauge similarity and embedding, and its time complexity is bounded by $O (|V|)$~\cite{pimentel2019efficient}. On the other hand, Node2Vec \cite{grover2016node2vec}(NV) uses a flexible biased random walks algorithm that combines breadth-first and depth-first sampling to generate node sequences while preserving homophily and structural equivalence. For a node, $u \in V$, a neighborhood $\Gamma_s (u)$ is defined as a set of nodes traversed by the surfer starting at $u$.  

\begin{figure}[h!]
\centering
\includegraphics[width=2.5in]{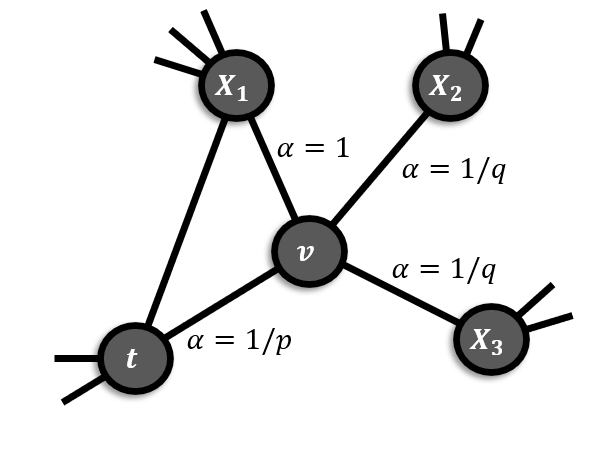}
\caption{Walk transitioned from node $t$ to $v$ and is now evaluating its next step. The transition probability is given by $\alpha$. The illustration has been redrawn from \cite{grover2016node2vec}. }
\label{fig:nv}
\end{figure}

\noindent Node2Vec modulates the exploration of neighborhood through parameters $p$ and $q$ (see Fig. \ref{fig:nv}). A low $p$ $(< \min(q, 1))$ enables the surfer to backtrack often and stay restricted to a local neighborhood (breadth-first search-like), whereas $q < 1$ makes the surfer explore a larger neighborhood (depth-first search-like). To generate embeddings, Node2Vec maximizes the log probability: 

    \begin{equation}
        \max_{\phi} \sum_{u \in V} \log \prod_{v \in \Gamma_s (u) } P ( v | \phi_u )
    \end{equation}

\noindent $P ( v | \phi_u )$ is the softmax function calculated using the embeddings of nodes $u$ and $v$, i.e., $\frac{\exp (\phi_u, \phi_v) }{\sum_w \exp (\phi_u, \phi_w)}$.

Evidently, both DeepWalk and Node2Vec capture the topological properties of nodes, whereby nodes that are located close to one another in the network tend to possess similar embeddings. However, another embedding model called \textit{struct2Vec}, represents the structural similarity of the nodes. where two nodes with similar local neighborhoods have similar embeddings even if they belong to different components of the network~\cite{ribeiro2017struc2vec}. To achieve this, it employs a hierarchical approach: similarity at the bottom of the hierarchy depends on the local neighborhoods, whereas at the top, similarity depends on a larger neighborhood around the nodes being embedded.

\subsubsection{Graph Neural Networks Methods}\label{sec:gnn}
These models learn embeddings by applying the principles of deep neural networks to graphical data. In a graph convolution network (GCN), the features of the node are passed to their neighbor via message-passing~\cite{kipf2016semi}. Later, the passed information is fused with existing features of the node using a predefined function (namely, sum, mean, or maximum, etc.). These aggregated features are then passed to fully connected neural networks to learn updated embeddings given specific objective functions. The inclusion of many convolutions reflects a deeper collected knowledge of a wide area of the graph into the final embedding of the nodes. The resulting embedding representation is used for link prediction purposes. 

\begin{figure}[h!]
  \centering
  \begin{subfigure}[b]{0.45\textwidth}
    \centering
    \includegraphics[width=\textwidth]{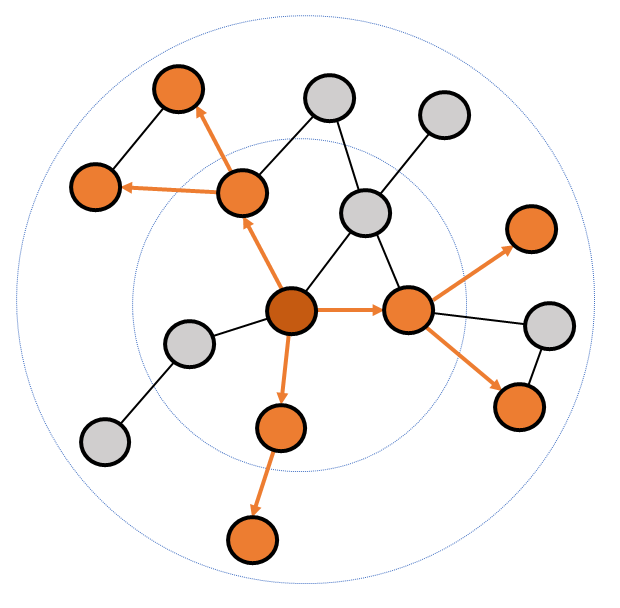}
    \caption{Sample neighborhood.}
  \end{subfigure}
  \hfill
  \begin{subfigure}[b]{0.45\textwidth}
    \centering
    \includegraphics[width=\textwidth]{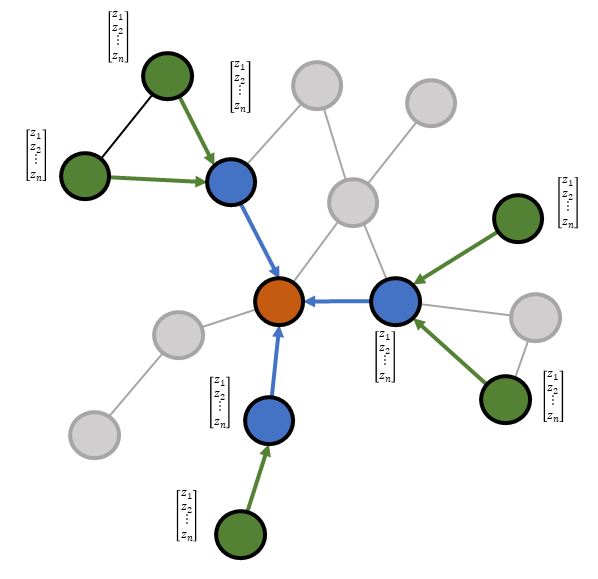}
    \caption{Aggregate feature information from neighbors.}
  \end{subfigure}
  \hfill
  \begin{subfigure}[b]{0.45\textwidth}
    \centering
    \includegraphics[width=\textwidth]{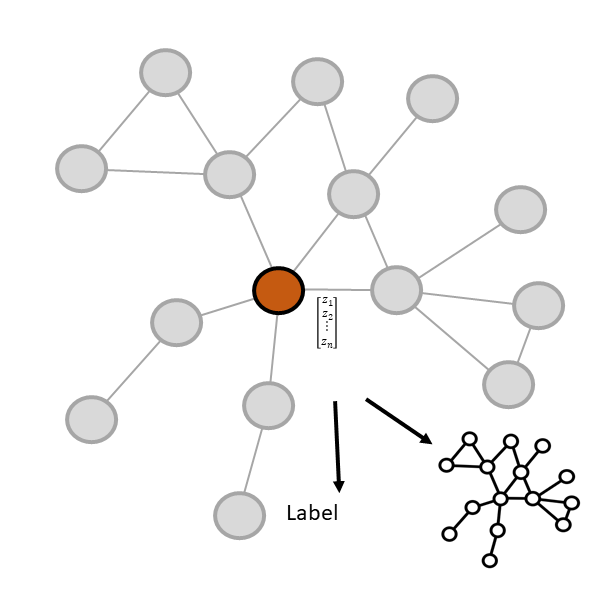}
    \caption{Predict graph context and label using aggregated information.}
  \end{subfigure}
  \caption{GraphSage workflow of 1) neighborhood sampling, 2) aggregating information from those samples, and then 3) prediction on the collected information. The illustration has been redrawn from \cite{hamilton2017inductive}.}
  \label{fig:sage}
\end{figure}

GraphSAGE, SAGE short for SAmple and aggreGatE, is a technique that combines sampling and aggregation of node-level features from the input graph~\cite{hamilton2017inductive}. As shown in Fig. \ref{fig:sage}, in the sampling phase, GraphSAGE gathers a neighborhood of depth $K$ around each node. In the aggregation phase, the embeddings of all the neighbors of node $v$, given by $\Gamma_v$, are aggregated into a single vector ($\mathbf{h}_{\Gamma_v}^k$), before concatenating the embedding of $v$ ($\mathbf{h}_v$) with the aggregated neighbor vector. Finally, the concatenated vector is passed through a fully connected layer with nonlinear activation function $\sigma$ controlled by weighted matrices $\mathbf{W}^k$, where $k = 1, 2, \cdots, K$. The stated steps are as follows: 

\begin{align}
    & \mathbf{h}_{\Gamma_v}^k = AGGREGATE_{k} ( \{ \mathbf{h}_u^{k - 1}, u \in \Gamma_v \} ) \; \\
    & \mathbf{h}_v ^k = \sigma (\mathbf{W}^k.CONCAT (\mathbf{h}_v ^{k-1}, \mathbf{h}_{\Gamma_v}^k))
\end{align}

\noindent The weight parameters ($\mathbf{W}^k$) are updated using a loss function $J$ that encourages neighboring nodes to get similar embeddings. Specifically, the function attempts to increase the similarity between the embeddings of $v$ and a node $u$ that co-occurs in random walks initiated at $v$ while reducing the similarity between $v$ and an arbitrary node $v_n$ sampled from a negative sampling distribution $P_n$, as: 

\begin{equation}
    J_G (v) = - \log ( \sigma (\mathbf{h}_v^T . \mathbf{h}_u) ) - Q. \mathbf{E}_{v_n \sim P_n (v)}  \log ( \sigma (- \mathbf{h}_v^T . \mathbf{h}_{v_n}) )  
\end{equation}

The other GNN is the graph autoencoder (GAE), which includes two components: encoder and decoder~\cite{kipf2016variational}. The encoder maps the nodes in the input network to their respective latent representations through a series of message-passing operations. The decoder, on the other hand, reads the latent factor and reconstructs the original graph structure. The overall training objective of a GAE is to minimize the error between the original graph and the reconstructed graph. During training, the encoder and decoder work in tandem, using backpropagation to minimize this loss. 

\subsection{Link Prediction in Temporal Networks}\label{sec:temporal}
Temporal networks are a sequence of many static networks, each marked with a timestamp $1, 2, \cdots, T$. As discussed in Sec. \ref{sec:type}, the purpose of link prediction (LP) metrics is to predict links at time $T + 1$ based on network connectivity till time $T$. We consider three temporal metrics that adapt existing static LP models, namely, collapsed tensor (CT)~\cite{liben2003link}, weighted collapsed tensor (WCT)~\cite{sharan2008temporal}, Jaccard, and non-negative matrix factorization (NMF). These metrics can be applied to unweighted networks with adjacency weights $A_{uv} = \{ 0, 1\}$ and weighted networks with $A_{uv} \in [0, 1]$. 

\begin{enumerate}

    \item In collapsed tensor, the scores for the edge weights at time $T+1$ ($S_{uv}^{T+1}$) are gauged to be the average of all edges across all prior timestamps $t = 1, 2, \cdots, T$, as shown in Eq. \ref{CT}. 
    
    \begin{equation}
        S_{uv}^{T+1} = \sum_{t=1}^T \frac{A_{uv}^t}{T}
        \label{CT}
    \end{equation}

    \item In weighted CT, the likelihood scores of edges at time $T+1$ ($S_{uv}^{T+1}$) are calculated as in Eq. \ref{WCT} where $\theta \in [0, 1]$ is a weighing parameter that assigns higher importance to recent links.
    
    \begin{equation}
        S_{uv}^{T+1} = \sum_{t=1}^T (1-\theta)^{T-t} A_{uv}^t
        \label{WCT}
    \end{equation}

    \item The temporal Jaccard coefficient leverages local similarity-based metrics (see Sec. \ref{sec:sim}), namely common neighbors, to predict future links at time $T + 1$ based on links at time $T$, as follows.
        
    \begin{equation}
        S_{uv}^{T + 1} = \frac{|N_u(W^T) \cap N_v(W^T)|}{|N_u(W^T) \cup N_v(W^T)|}
        \label{Jaccard}
    \end{equation}

    \begin{figure}[h!]
        \centering
        \includegraphics[width=5in]{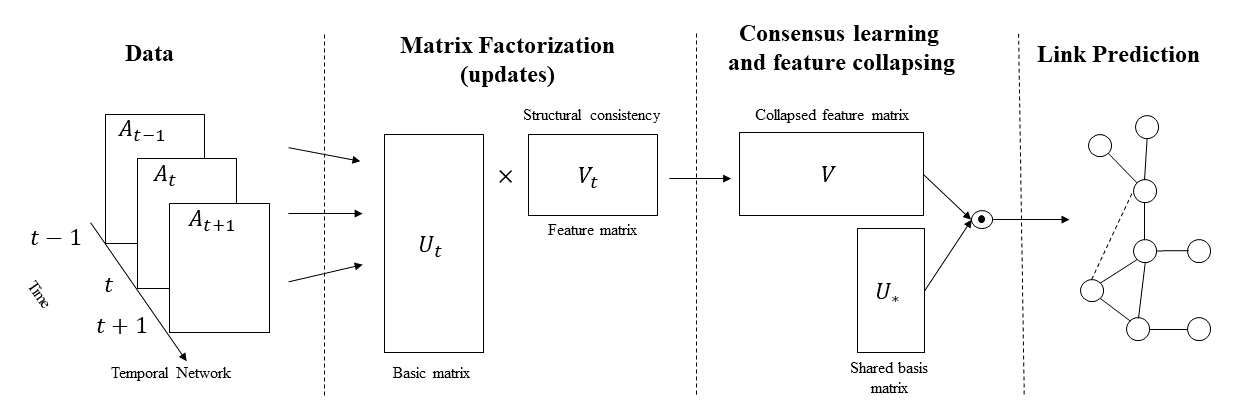}
        \caption{A schematic representation showing the steps in temporal non-negative matrix factorization-based link prediction. The matrices $U_t$ and $V_t$ are updated based on the adjacency matrices $A$ over three consecutive time instances $t - 1, t, t + 1$, before calculating the final matrices $V$ and $U_*$ to predict links in a network.}
        \label{fig:NMF}
    \end{figure}
    
    \item \textbf{Temporal LP using structural consistency regularized NMF}\cite{zhang2021temporal} is utilized to predict the network at time $T+1$. Here, the adjacency matrix $A_t$ is factorized into two non-negative matrices, the feature matrix $U$ and the coefficient matrix $V$. 
    An optimization function is applied to minimize the total difference between the current adjacency matrix $A_t$ and the product of its factorized matrices amounting to the low-rank approximation of $A_t$, i.e., $min(A_t-U_tV_t^T)$. This process is done by constantly updating the factor matrices over different time slots ($t$) to ensure the minimization objective function. All of the $U, V$, and $U_*$ are initialized with random values, and their final values are learned by adopting an iterative strategy that fixes two of the matrices while updating the third one (see Eqs. \ref{eq:update1} and \ref{eq:update2}). The final network at time $T+1$ can be calculated by Eqs. \ref{eq:update3} - \ref{eq:update5} following three phases: matrix factorization, consensus learning for basis matrices, and feature collapsing (see Fig. \ref{fig:NMF}). Note that $\alpha, \beta, \theta$ are weighing parameters with default values of $0.001$, $0.8$, and $0.8$, respectively.
    

    \begin{equation}\label{eq:update1}
        U_t = U_t \frac{\sum_{i=t-1}^{t+1} A_i V_t + U_*}{3 U_t V_t' V_t + U_t}
    \end{equation}

    \begin{equation}\label{eq:update2}
        V_t = V_t \frac{\sum_{i=t-1}^{t+1} A'_i U_t + \alpha A_i V_t}{3 V_t U_t' U_t + \alpha \sum_{i=t-1}^{t+1} D_i V_t}
    \end{equation}

    \begin{equation}\label{eq:update3}
        V = \sum _{t=1}^T \theta^{T-t} V_t
    \end{equation}

    \begin{equation}\label{eq:update4}
    U_* = \frac{1}{\sum_{i=1}^{t} \beta^{t-i}} \sum_{i=1}^{t} \beta^{t-i}U_i
    \end{equation}

    \begin{equation}\label{eq:update5}
        A_{T+1} = U_* V
    \end{equation}

\end{enumerate}
        


\section{Applications}\label{sec:apply}
\noindent In this section, we discuss the application of link prediction techniques in inferring disease-gene, protein-protein, disease-RNA, microbiome, drugs, and brain networks.

\subsection{Disease Gene Association Networks}\label{sec:app_dis_gene}

\begin{figure}[h!]
  \centering
  \begin{subfigure}[b]{0.6\textwidth}
    \centering
    \includegraphics[width=\textwidth]{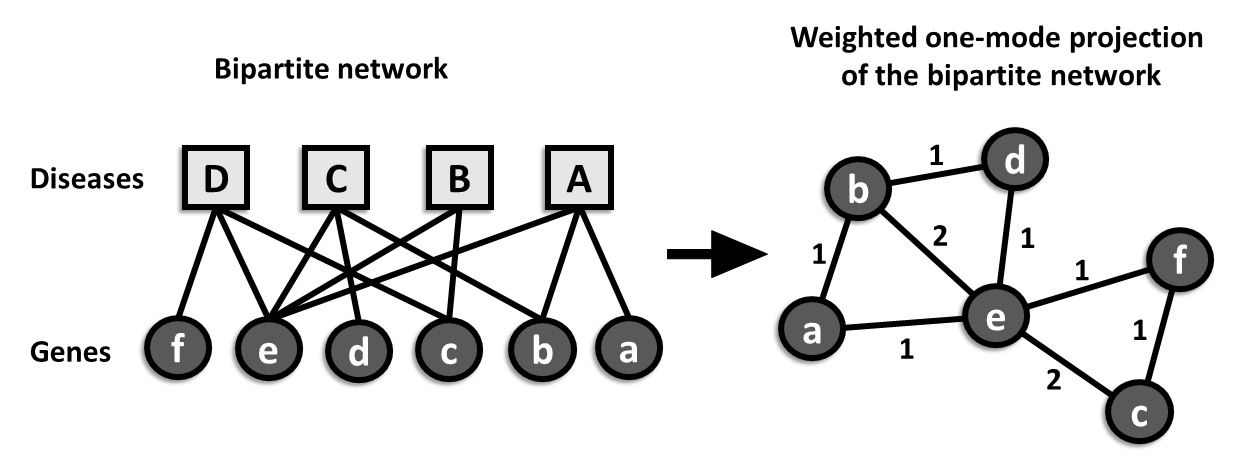}
    \caption{}
  \end{subfigure}
  \hfill
  \begin{subfigure}[b]{0.35\textwidth}
    \centering
    \includegraphics[width=\textwidth]{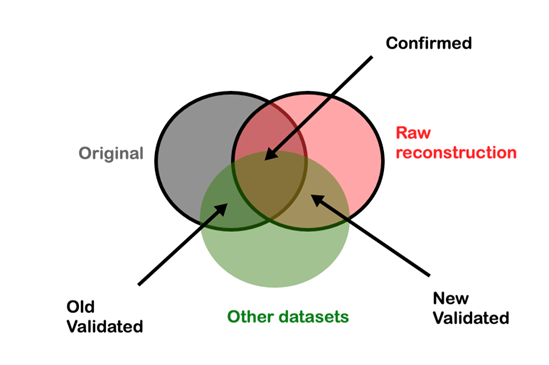}
    \caption{}
  \end{subfigure}
  \caption{Bipartite network of genes and diseases and its weighted one-mode projection. This illustration has been redrawn from \cite{alanis2014exploring}.}
  \label{fig:gene}
\end{figure}

Link prediction (LP) has been applied to gene networks, where the weights on links connecting two genes denote the number of shared attributes (such as diseases, drugs, ontology, etc.). Lobato et al. applied similarity-based LP measures on gene interaction networks, focusing on autoimmune diseases, namely, diabetic retinopathy, nephropathy, Kawasaki disease, systemic lupus erythematosus, celiac disease, rheumatoid arthritis ankylosing spondylitis, Crohn’s disease, primary sclerosing cholangitis, ulcerative colitis, type I diabetes, vitiligo, AIDS, hypothyroidism, and psoriasis~\cite{alanis2014exploring}. The gene networks were inferred from the bipartite gene-disease associations~\footnote{Bipartite networks are a class of networks comprising two groups of nodes (say, diseases and genes), where links may exist between nodes of different groups.} obtained from the \textit{genome-wide association studies} (GWAS) catalog~\cite{hindorff2009catalog}, by creating a new network where two genes are connected by a link (called an \textit{internal link}~\cite{allali2011link}) of weight equal to the number of common diseases they are associated with (see Fig. \ref{fig:gene}a). Their analysis employing weighted LP metric (see Table \ref{tab:local} for the formulation of weighted similarity-based measures) showed that weighted CN, AA, ACDD, and CAR can identify key gene associations for myriad autoimmune diseases.

\noindent Yang et al. leveraged LP to create a comprehensive network by integrating associations from 6 gene or protein association datasets~\cite{yang2017integration}. Their approach unfolded in the following three steps.

\begin{itemize}
    \item \textit{Raw networks.} For each network dataset, the combined likelihood score of a link between any pair of nodes was obtained as a weighted sum of scores from several LP metrics. Following this, the highest-scoring links are preserved in the 6 raw networks.

    \item \textit{Final reconstruction networks.} Each potential link in the raw network is classified into 3 groups: old, new, and confirmed. The old links are the ones in the original networks but not in the raw networks, while the new links are the ones present in the raw networks but not in the original networks. The confirmed links are the links that are present in both (see Fig. \ref{fig:gene}b). Finally, in each reconstruction network, the workflow preserves the confirmed links as well as the old or new links present in at least one of the other original networks. Each link in a reconstruction network is given a score equal to the weighted sum of its weight in the original network ($S_{FS}$) and raw network link weight ($S_{TS}$): 

        \begin{equation}
            S = \beta \times S_{FS} + (1 - \beta) \times S_{TS} \hspace{5mm} (\beta \in  [0, 1])
        \end{equation}
    
    \item \textit{Integrated network.} Given a score of a link in the $i^{th}$ final reconstruction network, $S_i$, the algorithm assigns a final link score given by: 
    
        \begin{equation}
            \mathbf{S} = 1 - \sqrt{ \prod_{i = 1}^{n} (1 - S_i)}
        \end{equation}    

\end{itemize}

\noindent As part of temporal link prediction of gene association in cancer, Zhang et al. \cite{zhang2021temporal} presented an improvement over conventional non-negative matrix factorization (refer to Sec. \ref{sec:MF}), termed \textit{structural consistency non-negative matrix factorization}, where the temporal adjacency matrices for networks of gene interactions $A$ is jointly decomposed into the corresponding basis and feature matrix representations ($B, F$) across three timepoints $t - 1, t, t + 1$ while preserving the intrinsic topological properties of the temporal networks in the feature vectors of the genes.

\subsection{Protein-Protein interaction networks (PPINs)} 

Protein-protein interaction networks are networks with proteins as nodes and links representing the interaction between pairs of proteins by means of structural and functional subunits called \textit{domains}. Kumar and Sharma proposed a metric that combines eigenvector centrality (see Sec. \ref{sec:central}) and the shortest path length to estimate the probability of a pair of nodes sharing a link. The proposed metric is defined as $S_{u,v}^SDEV=\frac{\sqrt{{evc}_u + {evc}_v}}{sp(u, v)}$, where ${evc}_u$ is the eigenvector centrality of node $u$ and $sp(u, v)$ is the length of the shortest path between nodes $u$ and $v$~\cite{kumar2021novel}.

The other LP approaches for PPINs involve graph embedding models (refer to Sec. \ref{sec:embed}). To incorporate both structural and functional attributes into association prediction, Nasiri et al. proposed an LP technique for unipartite as well as bipartite PPINs that also contain node weights capturing the functional attributes of the proteins. \textit{First}, Louvain community detection~\cite{blondel2008fast} is applied on a logical network of features to group similar features, before selecting a subset of features for the subsequent computations. \textit{Second}, a PPIN is constructed by connecting proteins based on a combination of similarities in topology and reduced features. \textit{Third}, DeepWalk (discussed in Sec. \ref{sec:random}) is applied to the PPINs to infer associations based on the Hadamard operator of the embedding vectors of a pair of nodes. Similarly, Kang et al. introduced a GNN-based model applied to PPINs from the STRING consortium~\cite{szklarczyk2015string}, where node embeddings were created using GCN encoder before employing a propagation rule to create link representations for predicting protein association~\cite{kang2022lr}. Finally, Zhao, C., et al. designed another GNN-based framework that incorporated two key features for the prediction of molecular interactions, namely, a mix-hop aggregator and contrastive self-supervised GNN~\cite{zhao2021csgnn}. While the mix-hop aggregator allows the updates to incorporate higher-order or indirect neighbor information into the node embeddings, contrastive self-supervised GNN preserves local and global network information to enhance the generalizability of the proposed model.

\subsection{Disease Non-coding RNA networks}
MicroRNAs and long non-coding RNAs affect gene expression controlling the onset and progression of several diseases, including rare and genetic ones~\cite{finotti2019micrornas}. Predicting the association between RNAs, namely, microRNA (miRNA) and long non-coding (lncRNA), and diseases is indispensable for comprehending the molecular underpinnings of diseases. The analysis of disease-RNA networks can help infer unknown relationships between the two entities, resulting in the identification of therapeutic intervention. Wen et. al constructed a bipartite network based on the known miRNA-disease associations, where each miRNA (or disease) is expressed as an association profile via a binary feature vector~\cite{zhang2019fast}. Two similarity matrices, namely the miRNA-miRNA similarity and disease-disease similarity, are computed via the Fast Linear Neighborhood Similarity measure and association profiles. Label propagation is applied separately to both, before using a weighted average to predict the miRNA-disease associations.

The other LP techniques for such networks are based on graph neural networks (GNNs), as discussed in Sec. \ref{sec:gnn}. Silva and Spinosa harnessed the graph autoencoder (GAE) framework to acquire the latent representation of nodes' features and edges and employed a neural network classifier to predict edges~\cite{silva2021graph}. Similarly, Shen et al. utilized GNN to predict ncRNA-protein interactions (NPI) across five different datasets. The model was developed by integrating GraphSAGE on a bipartite network of NPIs as input~\cite{shen2021npi}. Finally, Li et al. employed GNN for miRNA-disease associations. The model first constructs a bipartite graph of miRNAs and diseases to represent their associations. An encoder is then used to generate low-dimensional embeddings of miRNA and disease nodes, while consolidating heterogeneous information from their respective neighborhoods, using an aggregator function and a multi-layer perceptron. As a last step, the embeddings are fed into a bilinear decoder to detect potential connections between miRNA and disease nodes~\cite{li2021graph}.

\subsection{Microbiome Networks}

Research on the association between the imbalance in the gut microbial community (called \textit{dysbiosis}) and systemic diseases is at a nascent stage. Existing research aims to study the role of the oral microbiome in the development of rheumatoid arthritis, diabetes mellitus, and pancreatic cancer. The scientific community is relying on the advent of high-throughput sequencing, AI, and big data technologies, to establish the effect of microbes on health. It is important to study the confounding caused by non-modifiable factors (like age, gender, ethnicity, and genetic makeup) and modifiable factors (i.e., lifestyle, oral health status, and systemic factors)~\cite{seneviratne2020oral}. 

It is worth noting that the prediction of microbiome-disease associations has not always been formulated as a graph problem but as a supervised and deep machine-learning one. For instance, Larsen and Dai employed support vector machines (SVMs) to predict host status (dysbiotic or non-dysbiotic) from microbiome feature information, such as the microbiome community structures, predicted community enzyme function profiles, total and secondary community metabolomes~\cite{larsen2015metabolome}. Dahl et al. employed random forests (RFs) to study the effect of maternal gut microbiome on premature deaths from fecal samples of 121 mothers~\cite{dahl2017gut}. Reiman et al. employed CNNs to investigate how the microbial markers of the host may determine their microbiome interactions contributing to subsequent diseases~\cite{reiman2020popphy}. Metwally et al. utilized a long short-term memory approach to forecast food allergies during early life based on the longitudinal profiles of the gut microbiome in subjects, exhibiting improved predicted power than RFs, SVMs, and deep neural networks~\cite{metwally2019utilizing}.

Finally, Fu et al. leveraged a natural language processing technique to generate a knowledge graph (KG) of microbes and diseases~\cite{fu2020integrated}. Next, they further enriched the KG by incorporating additional bacterial feature information, such as strain, salinity, oxygen requirement, temperature range of growth environment, and habitat, through clustering and association analysis. Lastly, they trained deep learning, matrix factorization, and KG embedding models on triplets of the head entity, relationship, and tail entity of the KG to infer microbe-disease associations. 

\subsection{Drug Association}

Drug association networks capture the relationship between potential drug targets for diseases, the mechanism of action of a drug on a disease, or the side effects of a given drug. The simplest representation of drug associations is a bipartite network, where links may only exist between different entity types, like drugs and side effects, drugs and diseases, etc. Luo attempted to infer the toxicity of drugs through local similarity-based LP metrics (refer to Sec. \ref{sec:local}) on a bipartite network of diseases and their side effects~\cite{luo2014predicting} generated from the SIDER2 dataset containing side effect frequency for a drug~\cite{kuhn2010side}. Gundogan et al.~\cite{gundougan2017link} applied the similarity-based metrics to a disease-drug bipartite network inferred from a database of drugs, diseases, percentage of users who prefer these drugs, class of drugs, etc.~\cite{drugs2006drugs}, in a manner depicted in Fig. \ref{fig:gene}a.

\begin{figure}[h!]
\centering
\includegraphics[width=5in]{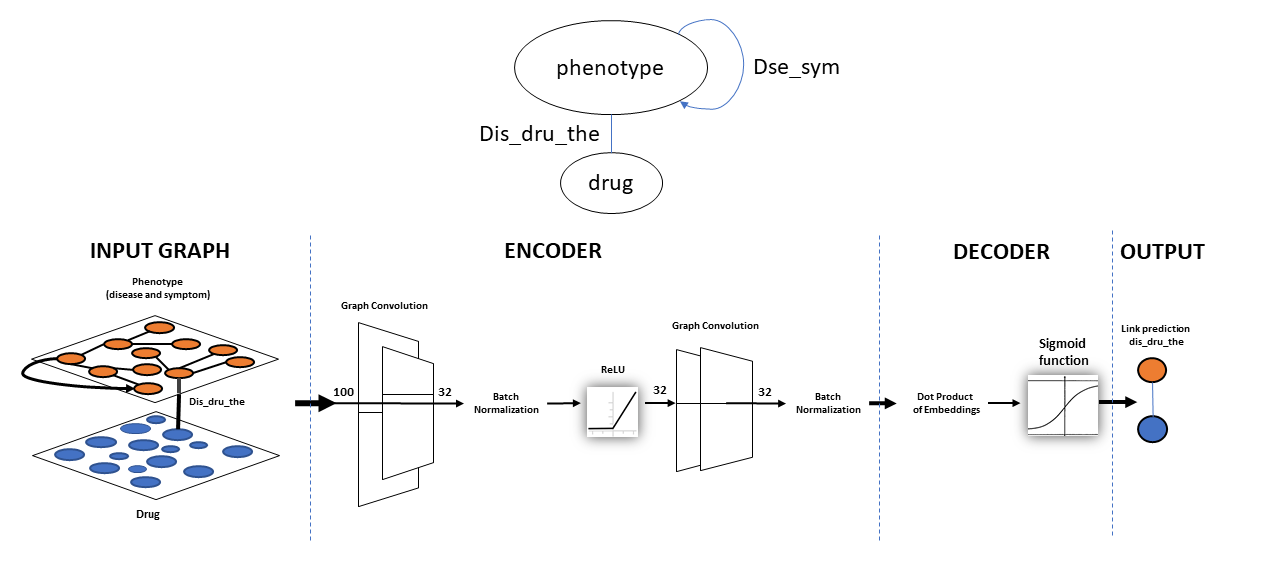}
\caption{REDIRECTION architecture showing a drug-phenotype interaction (top) and the phases of encoder-decoder model architecture (bottom). The figure has been redrawn from \cite{griesser2008redirection}.}
\label{fig:drug_app}
\end{figure}

Others employed graph neural network (GNN) based methods for predicting drug-disease associations. A key application of LP lies in \textit{drug repurposing}, where the goal is to find therapeutic utilities for existing drugs. Munoz et al. designed a GNN-based workflow, called REDIRECTION~\cite{griesser2008redirection}, to find new drug-disease associations from a biomedical knowledgebase, called DISNET~\cite{prieto2022disnet}, containing information regarding diseases, symptoms, and drugs. REDIRECTION is the drug-disease network (see Fig. \ref{fig:drug_app} top) as input and applies a two-layer convolution approach, each containing normalization and nonlinearity activation function (ReLU). The embeddings are encoded using GraphSAGE (see Sec. \ref{sec:gnn}). Next, the decoder calculates the dot product of embedding followed by sigmoid transformation to get the link score between a disease-gene pair (see Fig. \ref{fig:drug_app} bottom). Finally, the model parameters are optimized using the Binary Cross Entropy as the loss function. Similarly, Wang et al. extended a graph convolutional network model, Decagon, on the identification of potential drug targets~\cite{wang2021drug}. Instead of leveraging the bipartite relationship, they used open-source drug datasets, namely, Drugbank, SIDER, HPRD, etc., to identify target-target, drug-target, and drug-drug relationships as links, while drug and target characteristics were incorporated as node attributes and exhibited high prediction accuracy on the drug-disease heterogeneous network. 

\begin{figure}[h!]
  \centering
  \begin{subfigure}[b]{0.45\textwidth}
    \centering
    \includegraphics[width=\textwidth]{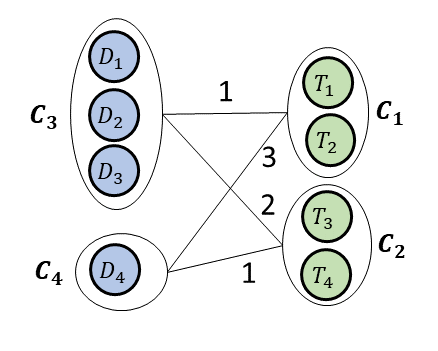}
    \caption{}
  \end{subfigure}
  \hfill
  \begin{subfigure}[b]{0.35\textwidth}
    \centering
    \includegraphics[width=\textwidth]{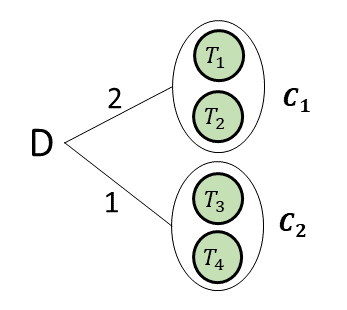}
    \caption{}
  \end{subfigure}
\caption{Community to community matching example. The labels assigned to the edges indicate the count of connections between vertices belonging to matched communities. This illustration has been redrawn from \cite{koptelov2021lpbycd}.}
\label{fig:com}
\end{figure}

The other LP approach for drug interaction prediction involves community detection and supervised machine learning (ML). \textit{First}, given a multilayer bipartite network of drugs and their targets with links existing between a pair of drug nodes, a pair of target nodes, and a drug-target pair, an approach by Koptelov et al. predicts unknown drug-target associations~\cite{koptelov2021lpbycd}. To this end, pure communities (T) comprising only drugs (D) and only targets (T) are first identified within the network, before applying local similarity-based LP metrics in two ways: community-to-community and node-to-community. In community-to-community network representation, LP is applied to a network of communities as nodes and links exist between pure drug community and target communities  (see Fig. \ref{fig:com}a); whereas in node-to-community, links exist between a drug community and a target community, and vice versa (Fig. \ref{fig:com}b). In both, the links are weighted, denoting the number of links existing between community pairs and community-node pairs in the original drug-target bipartite network. \textit{Second}, Jiang et al. proposed a two-step approach, where embeddings of diseases and drugs were learned from a network of drugs, diseases, proteins, miRNAs, and lncRNAs before using a supervised ML (random forest model) to predict drug-disease associations. The proposed approach was verified on several benchmark multi-biomolecular networks as well as by successfully predicting a ranked list of genes connected to two common human diseases~\cite{jiang2022effective}.

\subsection{Brain networks} 

Brain networks capture the interaction among different regions of the brain. The latest imaging technologies made it possible to model the significant changes in the anatomy and connections in the brain caused by disease conditions. Efforts have been made to leverage the known topological attributes of brain connectome, such as degree, clustering coefficients, the shortest path length, transitivity, efficiency, etc., in combination with the local link prediction (LP) approach (see Sec. \ref{sec:local}) to track and predict these changes in connectivity~\cite{he2018link}. He et al. considered the time-varying \textit{electroencephalography} (EEG) dataset of epilepsy patients during seizures before using a measure called, \textit{Phase Locking Value} (PLV), to infer the strength or weight of connections among different parts of the brain. Their analysis showed that resource allocation (RA) improved greater link prediction accuracy on the weighted networks than other local LP measures, namely, common neighbor (CN), Adamic-Adar (AA), and Sorenson algorithms.

\begin{figure}[h!]
  \centering
  \begin{subfigure}[b]{0.45\textwidth}
    \centering
    \includegraphics[width=\textwidth]{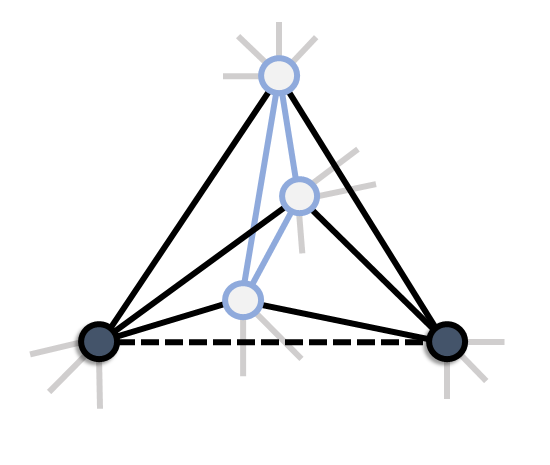}
    \caption{}
  \end{subfigure}
  \hfill
  \begin{subfigure}[b]{0.45\textwidth}
    \centering
    \includegraphics[width=\textwidth]{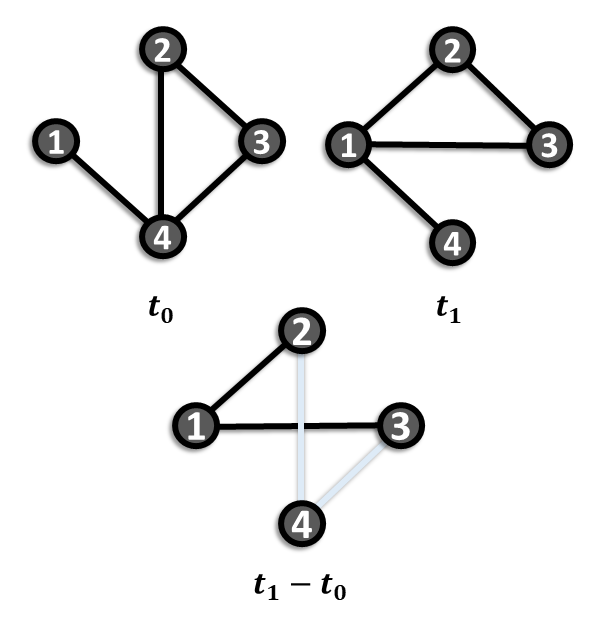}
    \caption{}
  \end{subfigure}
\caption{Prediction in brain networks: (a) Cannistraci-Alanis-Ravasi (CAR) where two nodes (colored black) are likely to be connected if their immediate neighbors (colored white) are mutually connected; (b) differential connectivity in brain network, where links (colored white) present at time $t_0$ disappear at time $t_1$ and new links (colored black) emerge at time $t_1$.}
\label{fig:brain}
\end{figure}

Cannistraci et al. compared the performance of local metrics against a stack of path-based approaches, called the \textit{Cannistraci-Alanis-Ravasi} (CAR) index (refer to Sec. \ref{sec:path}), which proposes that two nodes are more likely to be linked if their immediate neighbors are mutually connected forming a small community~\cite{cannistraci2013link}. Their analysis of two networks (cortical connectome and frontal ganglia connectome) with neurons as nodes and synaptic connections as links show that the combination of CAR and preferential attachment, \textit{Cannistraci preferential attachment} (CPA) achieves the highest prediction accuracy. The authors intuit that the formation of new synapses during the learning process could be a function of the existing local synaptic communities (see Fig. \ref{fig:brain}a). 

Sulaimani et al. applied local LP metrics to predict the evolution (i.e., addition and removal) of links in brain networks of Alzheimer’s disease~\cite{sulaimany2017predicting}. Specifically, given a network acquired by whole-brain \textit{magnetic resonance imaging} (MRI) of the patients at time $t$, the goal of their time-varying LP problem is to predict the changes in connectivity at $t^\prime$ ($t^\prime > t$). At any given $t$, a differential matrix is created by subtracting the adjacency matrix at $t - 1$ from that of $t$. Thus, an entry of  0, 1, and -1 denote no change, link addition, and link removal, respectively (refer to Fig. \ref{fig:brain}b for a schematic of the differential matrix). While addition is predicted using the high likelihood of association based on a metric, link removal is achieved by flipping the bits of the matrix before applying LP. This study also utilizes local LP metrics, namely, RA, CN, AA, PA, and Jaccard, to show that AA can track the changes in connectivity at different stages of Alzheimer’s progression. 

\section{Performance Evaluation}

To assess the effectiveness of various static as well as dynamic LP models (see Sec. \ref{sec:method}), we select representative models from the similarity, centrality, embedding, and temporal categories and evaluate their performance using specific datasets. We conduct a comprehensive evaluation of the models using a 5-fold cross-validation approach and report the mean (and standard deviation of) the area under the curve (AUC) for the receiver operating characteristic (ROC) across the folds.

\subsection{Evaluation Techniques and Datasets}\label{sec:eval}

To predict links in static biological networks (see Table. \ref{tab:network-stats}), we used \textit{CAR}, \textit{CH2\_L2} and \textit{CH2\_L3} (local and global similarity-based metrics); \textit{DeepWalk}, \textit{Node2Vec}, \textit{GCN}, \textit{graphSAGE} and \textit{GAE} (embedding and neural network methods); and \textit{CCPA} and \textit{ECC} (centrality-based methods). Each input biological network $G = (V, E)$ is split into training and testing graphs $G^{\tau}$ and $G^P$, containing 80\% and 20\% of the edges in $E$, respectively (details discussed in Sec. \ref{sec:setting}). Similarly, to predict links at time $T + 1$ in temporal networks ($G_{T + 1}$), the training set comprises the network snapshots of earlier time instances, i.e., ($G_1, G_2, \cdots, G_T$). We used the following parameters: learning rate $0.01$, 100 training epochs, node embedding size $8$, and 32 randomly chosen embedding features. For the random walk-based models, the number and the length of random walks are both set to $100$. For supervised models, the nodes in $G^{\tau}$ are labeled using the Louvain community detection algorithm~\cite{de2011generalized}. 




\begin{table}[ht]
\caption{Network Statistics, $|V|$: number of nodes, $|E|$: number of edges, GCC: transitivity, ACC: Avg. clustering coeff., $D$: Density, $r$: Assortativity, ASP: Avg. shortest path, $d$: diameter, MOD: modularity. (The abbreviation \textit{ts.} against the dynamic networks DPPIN refers to the number of timestamps or network snapshots.)}
\label{tab:network-stats}
\centering
\begin{tabular}{llccccccccc}
\hline
Networks	&	$|V|$	&	$|E|$	&	GCC	&	ACC	&	$D$	&	$r$	&	ASP	&	$d$	&	MOD	\\
\hline
ENZYMES\_g296 \cite{nr}	&	125	&	282	&	0.029	&	0.006	&	0.018	&	0.287	&	12.94	&	32	&	0.758	\\
bn-mouse\_visual-cortex\_2 \cite{nr}		&	193	&	428	&	0.005	&	0.021	&	0.012	&	-0.845	&	4.271	&	8	&	0.753	\\
bn-macaque-rhesus\_brain\_1	 \cite{nr}	&	242	&	6108	&	0.337	&	0.450	&	0.105	&	-0.055	&	2.218	&	4	&	0.307	\\
bio-yeast-protein-inter	 \cite{nr}	&	1870	&	4480	&	0.055	&	0.067	&	0.001	&	-0.156	&	6.812	&	19	&	0.847	\\
bio-grid-mouse	 \cite{nr}	&	2900	&	6544	&	0	&	0	&	0.0008	&	-0.153	&	9.555	&	31	&	0.925	\\
bio-celegans-dir \cite{nr}		&	453	&	4065	&	0.124	&	0.647	&	0.020	&	-0.22	&	2.664	&	7	&	0.405	\\
Se-DoDecagon\_sidefx \cite{biosnapnets}	&	594	&	1118	&	0	&	0	&	0.0032	&	-0.526	&	1.978	&	2	&	0.926	\\
PDN	&	437	&	5753	&	0.119	&	0.168	&	0.030	&	-0.131	&	0	&	0	&	0.329	\\
\hline
DPPIN-Babu~\cite{fu2022dppin} (36 ts.) & 5003 & 111,466  &		&	&	&	&	&	&	\\
DPPIN-Breitkreutz~\cite{fu2022dppin} (36 ts.) & 869 & 39,250  &		&	&	&	&	&	&	\\
DPPIN-Yu~\cite{fu2022dppin} (36 ts.) & 1163 & 3602  &		&	&	&	&	&	&	\\
\hline
\end{tabular}
\end{table}

\subsection{Performance Evaluation of the static models}
\begin{figure}[h!]
  \centering
  \begin{subfigure}[b]{0.45\textwidth}
    \centering
    \includegraphics[width=\textwidth]{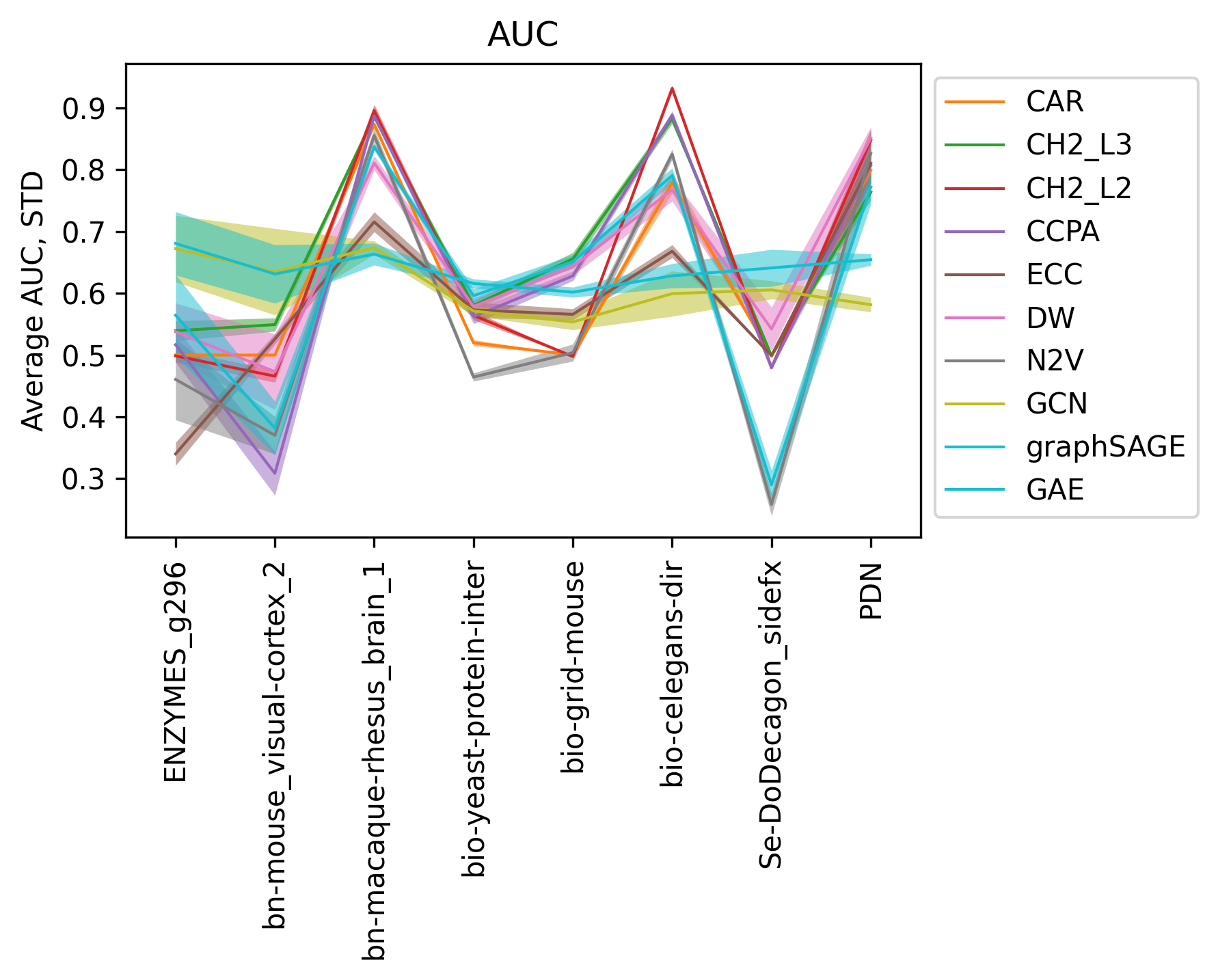}
    \caption{AUC (STD) performance.}
  \end{subfigure}
  \hfill
  \begin{subfigure}[b]{0.45\textwidth}
    \centering
    \includegraphics[width=\textwidth]{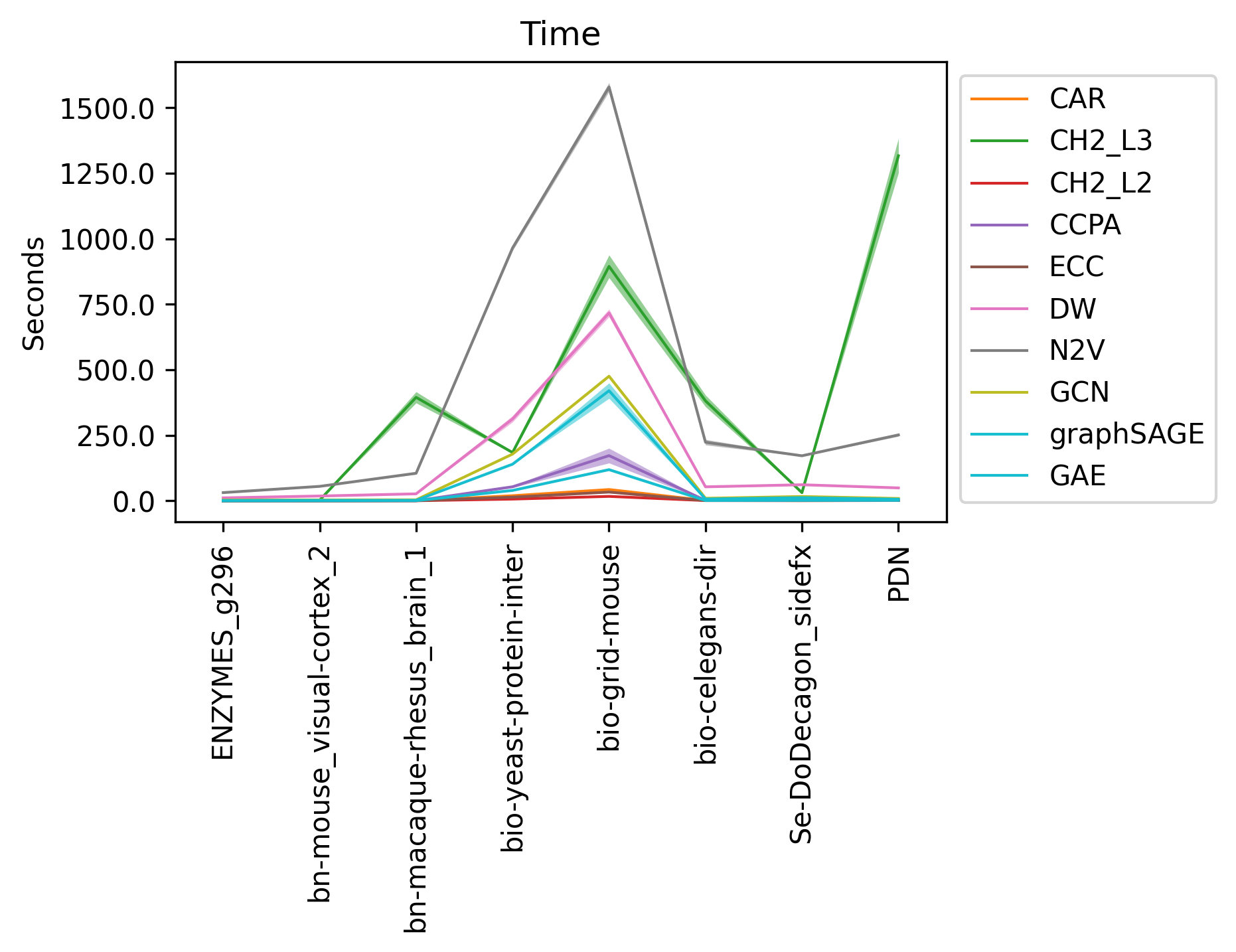}
    \caption{Time consumption in seconds.}
  \end{subfigure}
  \caption{Performance evaluation and time consumption of the ten different models implemented on different networks. We reported the average AUC values and their standard deviation using 5-fold cross-validation. The Models are CAR, CH2\_L3, CH2\_L2, CCPA, ECC, DW, N2V, GCN, GraphSAGE and GAE.}
  \label{fig:fig8}
\end{figure}

Fig \ref{fig:fig8}a shows that there is considerable variation in the performance of different LP models, in terms of their AUC values, across different networks. 

Cannistraci-Alanis-Ravasi-based variation of the resource allocation (CAR) presents consistently average performance (AUC $0.5 - 0.7$) across all networks, indicating a reliable but not leading efficacy. CH2\_L2 varies greatly, leading in \textit{bn-macaque-rhesus\_brain\_1} and \textit{bio-celegans-dir}, yet falling to lower ranks in networks like \textit{ENZYMES\_g296} and \textit{bio-grid-mouse}, showing a network-dependent performance. CH2\_L3, similar to CH2\_L2, excels in \textit{bn-macaque-rhesus\_brain\_1} and \textit{bio-celegans-dir}, showing particular strength in certain biological networks.

Both centrality-based parameterized algorithm (CCPA) and edge clustering coefficient (ECC) consistently show average AUC performance across networks, with their closest performance observed in \textit{bio-celegans-dir}. The similarity in the performance characteristics of ECC and CCPA can be ascribed to their emphasis on the immediate relationships and interactions between nodes. Recall from Sec. \ref{sec:central}, ECC-based link prediction evaluates the probability of edge formation by considering the degree of clustering around an edge, emphasizing local connectivity. Likewise, CCPA, incorporating node closeness and common neighbors, leverages proximity and direct ties.

DeepWalk performs notably well in PDN, suggesting it is well-suited for networks with a structure that aligns with random walk-based methods, while Node2Vec's performance is more varied but tends to follow DW closely, indicating it may share some methodological strengths and weaknesses. Graph convolutional network (GCN) and GraphSAGE are the top performers in several networks, with GraphSAGE taking the lead in \textit{ENZYMES\_g296} and \textit{Se-DoDecagon\_sidefx}, showcasing the potency of graph neural network models in these domains. Graph autoencoder models, while sharing the GNN approach, show more variability, suggesting that the nuances of GNN architecture and parameters can significantly impact performance. Overall, the GNN models (GCN, GraphSAGE, GAE) tend to rank higher, reflecting their sophisticated representation learning capabilities.

Fig. \ref{fig:fig8}b shows the average time consumed in seconds for 5-fold cross-validation experiments. N2V, CH2\_L3, and DW exhibit the longest running times. This extended runtime is attributed to the comprehensive exploration of a wider neighborhood during model training. In contrast, CH2\_L2 and CAR have the shortest runtimes, as they confine their exploration to a smaller neighborhood.


\subsection{Comparison of Link Prediction Models}

\subsubsection{Exploration versus Efficiency}\label{sec:tradeoff}

Neighborhood exploration during the training phase determines the efficacy of a model in predicting links in biological networks. The local LP metrics (see Sec. \ref{sec:local}), which rely on graph similarity, offer computational efficiency as they investigate a limited neighborhood around each node. The focus on proximity in terms of common neighborhoods between pairs of nodes may often lead to a narrower scope that can miss critical long-range relationships \cite{mayo}, potentially sacrificing prediction accuracy. On the other hand, centrality and embedding-based LP models (see Secs. \ref{sec:central} and \ref{sec:embed}) encompass a broader neighborhood. They gather information from multiple hops, aggregating relationships over extended network distances to formulate their predictions. This broader perspective, though more computationally intensive, often captures richer network patterns of intricate associations among biological entities.

The exploration of a very large neighborhood, while beneficial in capturing long-range relationships, can present its own challenges. For instance, when the neighborhood is expansive, the model may generate embeddings that are excessively similar and indistinguishable, limiting its discriminative power. This over-smoothing phenomenon \cite{li2018deeper} can result in a network representation where nodes exhibit little diversity in their predictive features. Thus, to determine the ideal LP model, one must weigh these trade-offs. Researchers must consider the balance between computational efficiency and prediction accuracy in their respective domains. The key lies in tuning the model parameters to strike a balance in neighborhood exploration, ensuring that the LP model can provide optimal predictions while navigating the heterogeneous biological and biomedical networks. This delicate calibration of neighborhood size, guided by the unique characteristics of the biological system, is pivotal to achieving the best predictive performance.

\begin{figure}[h!]
  \centering
  \begin{subfigure}[b]{0.4\textwidth}
    \centering
    \includegraphics[width=\textwidth]{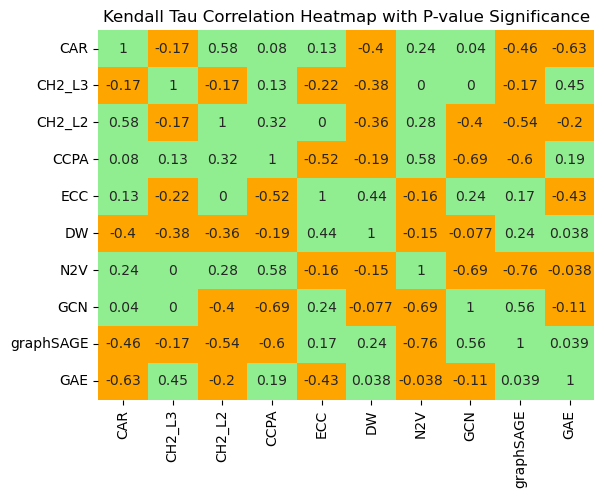}
    \caption{}
  \end{subfigure}
  \hfill
  \begin{subfigure}[b]{0.55\textwidth}
    \centering
    \includegraphics[width=\textwidth]{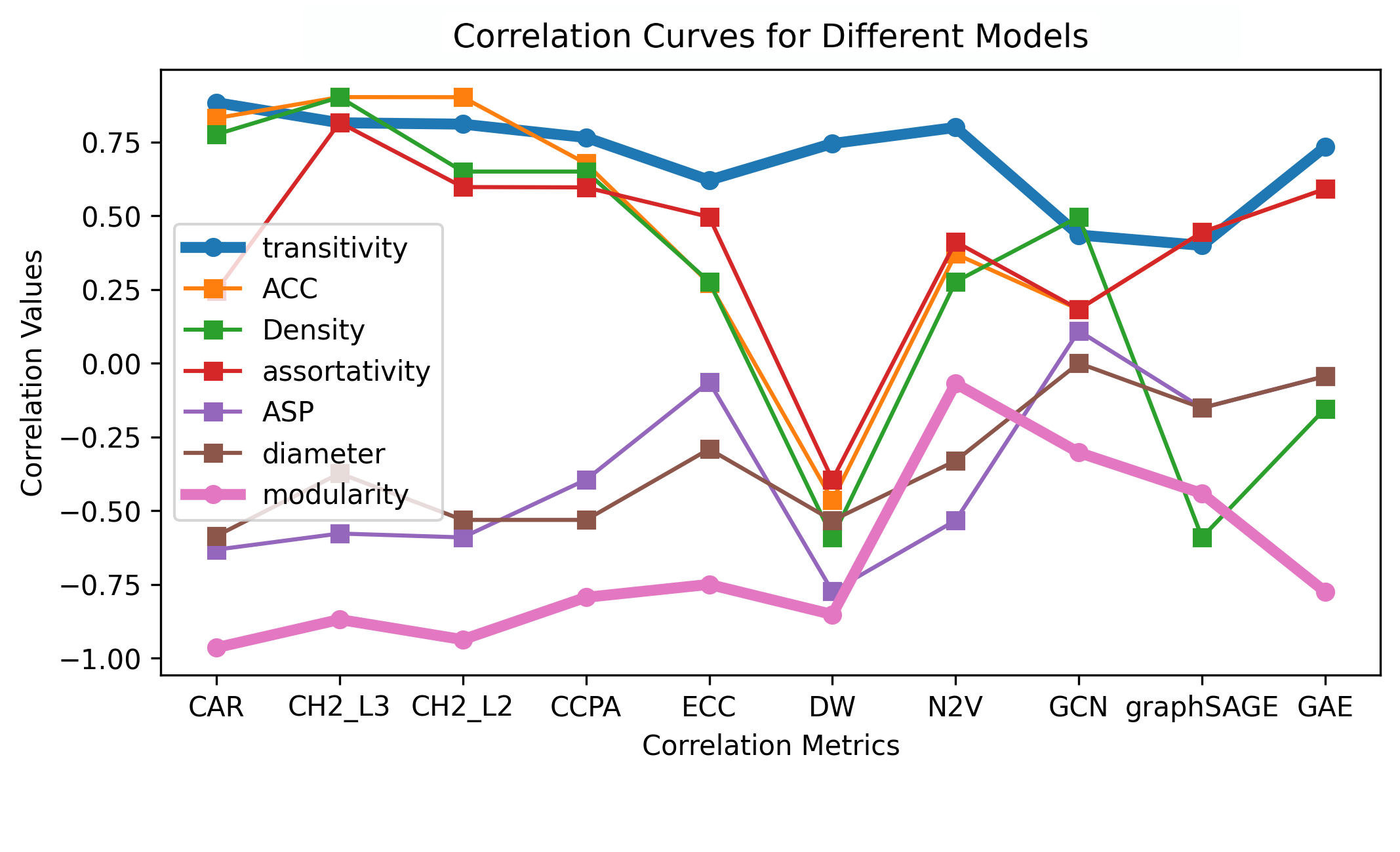}
    \caption{}
  \end{subfigure}
  \caption{Comparison of the performance accuracy of the LP models: (a) Kendall Tau correlation heatmap with p-value significance. cells are colored orange if their p-value is greater than 0.05, otherwise, it is colored light green; (b) Pearson correlation coefficient between the AUC and values for a given topological property across all the biological network datasets.}
  \label{fig:heat}
\end{figure}

We compare the extent of correlation in the performance of the LP models. For each biological network, we rank the AUC scores from different network models from best to worst (see Table \ref{tab:rank_rows_static}). Following this, we calculate the \textit{Kendall rank correlation coefficient} between each pair of LP models, where the corresponding scores represent the extent of their similarity. Fig. \ref{fig:heat}a shows the coefficients between each pair of models in green and red if the corresponding p-value is less than or equal and greater than $0.05$, respectively. It shows a lack of consensus among the models, suggesting that different models work well for different networks. However, there exists some correlation between the local and path-based models (namely, CAR, CH2\_L2, CH2\_L3, CCPA, etc) as well as representation learning models (i.e., GCN, GAE, and GraphSAGE)

\begin{table}[h]
\centering
\caption{Ranking of link prediction models across various networks. $L3, L2$ and $GS$ stands for $CH2\_L3, CH2\_L3$ and $graphSAGE$ respectively.}
\label{tab:rank_rows_static}
\begin{tabular}{l|ccc|cc|cc|ccc}
\hline
Networks & CAR & L3 & L2 & CCPA & ECC & DW & N2V & GCN & GS & GAE \\ \hline
ENZYMES\_g296 & 7 & 4 & 8 & 6 & 10 & 5 & 9 & 2 & 1 & 3 \\
bn-mouse\_visual-cortex\_2 & 5 & 3 & 7 & 10 & 4 & 6 & 9 & 1 & 2 & 8 \\
bn-macaque-rhesus\_brain\_1 & 4 & 3 & 1 & 2 & 8 & 7 & 5 & 9 & 10 & 6 \\
bio-yeast-protein-inter & 9 & 3 & 7 & 8 & 5 & 4 & 10 & 6 & 1 & 2 \\
bio-grid-mouse & 9 & 1 & 10 & 4 & 6 & 3 & 8 & 7 & 5 & 2 \\
bio-celegans-dir & 6 & 3 & 1 & 2 & 8 & 7 & 4 & 10 & 9 & 5 \\
Se-DoDecagon\_sidefx & 6 & 5 & 7 & 8 & 4 & 3 & 10 & 2 & 1 & 9 \\
\hline
\end{tabular}
\end{table}

Next, we scrutinize the impact of topological properties, as outlined in Table \ref{tab:network-stats}, on network performance accuracy. To achieve this, we compute the Pearson correlation coefficient between the Area Under the Curve (AUC) scores across all biological network datasets and their corresponding values for a given topological property. A high or low correlation signifies the potential of the associated topological property to assist link prediction models in identifying patterns.

The analysis result (see Fig. \ref{fig:heat}b) consistently reveals a robust correlation, denoted in bold blue, for network \textit{transitivity} (measured as the fraction of open triads that form triangles). This finding suggests that high transitivity corresponds to the existence of tightly interconnected communities or cliques. This structural coherence facilitates the identification of recurring patterns and relationships, enhancing the ability of link prediction models to discern and predict missing links. Conversely, network \textit{modularity}, denoted in bold magenta, exhibits an inverse correlation with high AUC scores, especially for similarity and path-based link prediction models. This is attributed to networks with low modularity, characterized by homogeneous communities where nodes share similar connectivity patterns. In such cases, local and random-walk-based link prediction models excel, leveraging the uniformity of local structural information to predict links effectively.

\subsubsection{Strengths and Weaknesses}\label{sec:comp} 

As discussed in Section~\ref{sec:tradeoff}, the exploration of a broader neighborhood has the advantage of effectively incorporating global patterns into our models. Notably, conventional models like CAR and CH2-L2 tend to exclusively consider one-hop neighborhoods, leading to considerably faster execution times but lacking the knowledge of global network information. In contrast, CH2-L3 presently explores multi-hop paths, which necessitates a more substantial computational effort. Similarly, models based on random walks can achieve accurate predictions through meticulous training but at the cost of increased computational overheads. Our analysis demonstrates that a random walk-based model, specifically DeepWalk and Node2Vec, trained using 100 random walks of length 100 starting at each node, attains a comparable performance accuracy. Lastly, when it comes to graph neural network (GNN)--based Link Prediction (LP) models, their success is contingent on the accuracy of the ground truth labels. In our analysis, we employed the Louvain community detection algorithm~\cite{de2011generalized} to assign cluster IDs to nodes. It is worth noting that the Louvain-based cluster labels may not always represent the optimal labeling for the nodes, which accounts for the below-par performance of the GNN-based models on some biological datasets.

\begin{figure}[h!]
  \centering
  \begin{subfigure}[b]{0.45\textwidth}
    \centering
    \includegraphics[width=\textwidth]{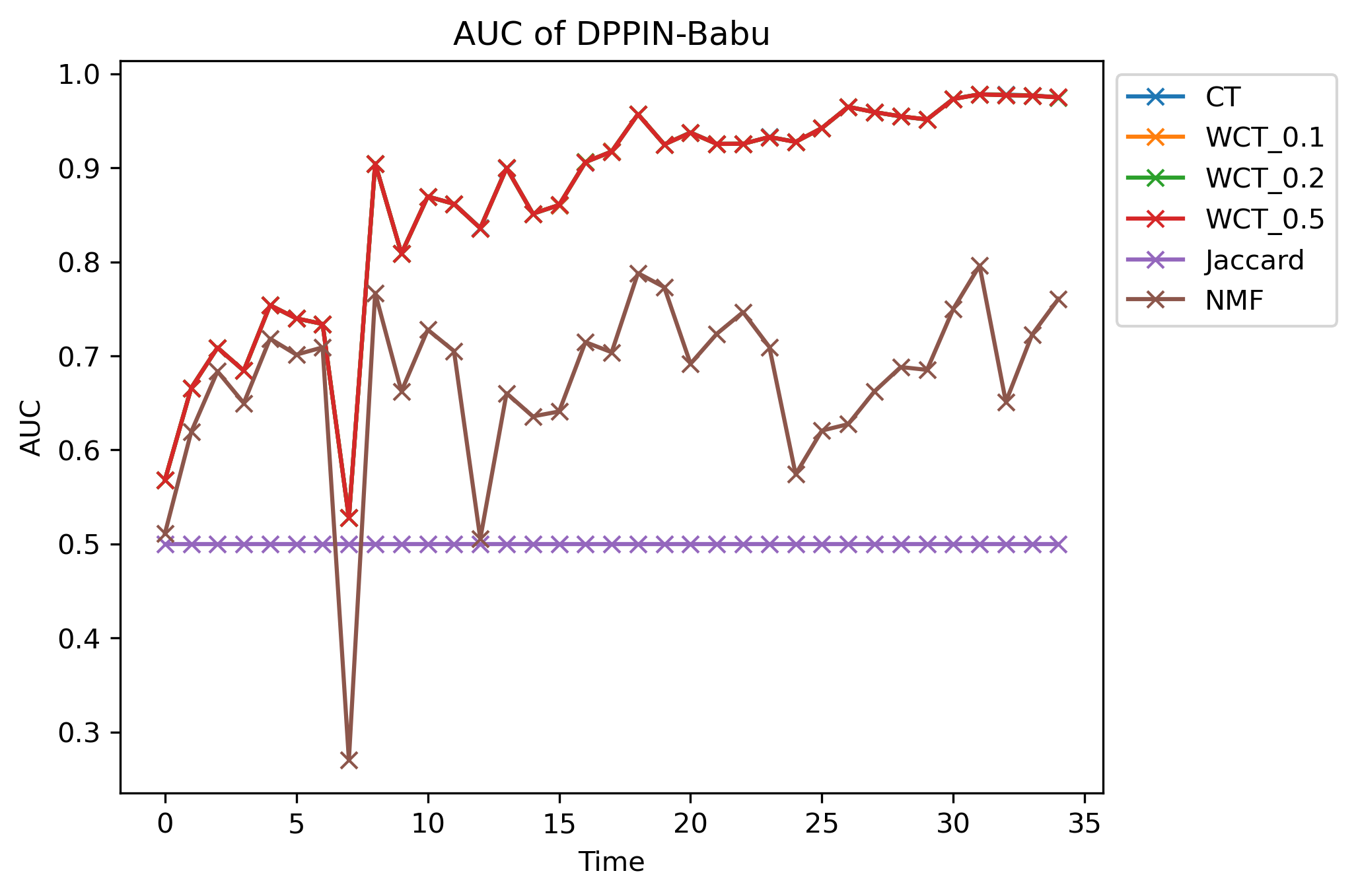}
    \caption{AUC performance of DPPIN-Babu network.}
  \end{subfigure}
  \hfill
  \begin{subfigure}[b]{0.45\textwidth}
    \centering
    \includegraphics[width=\textwidth]{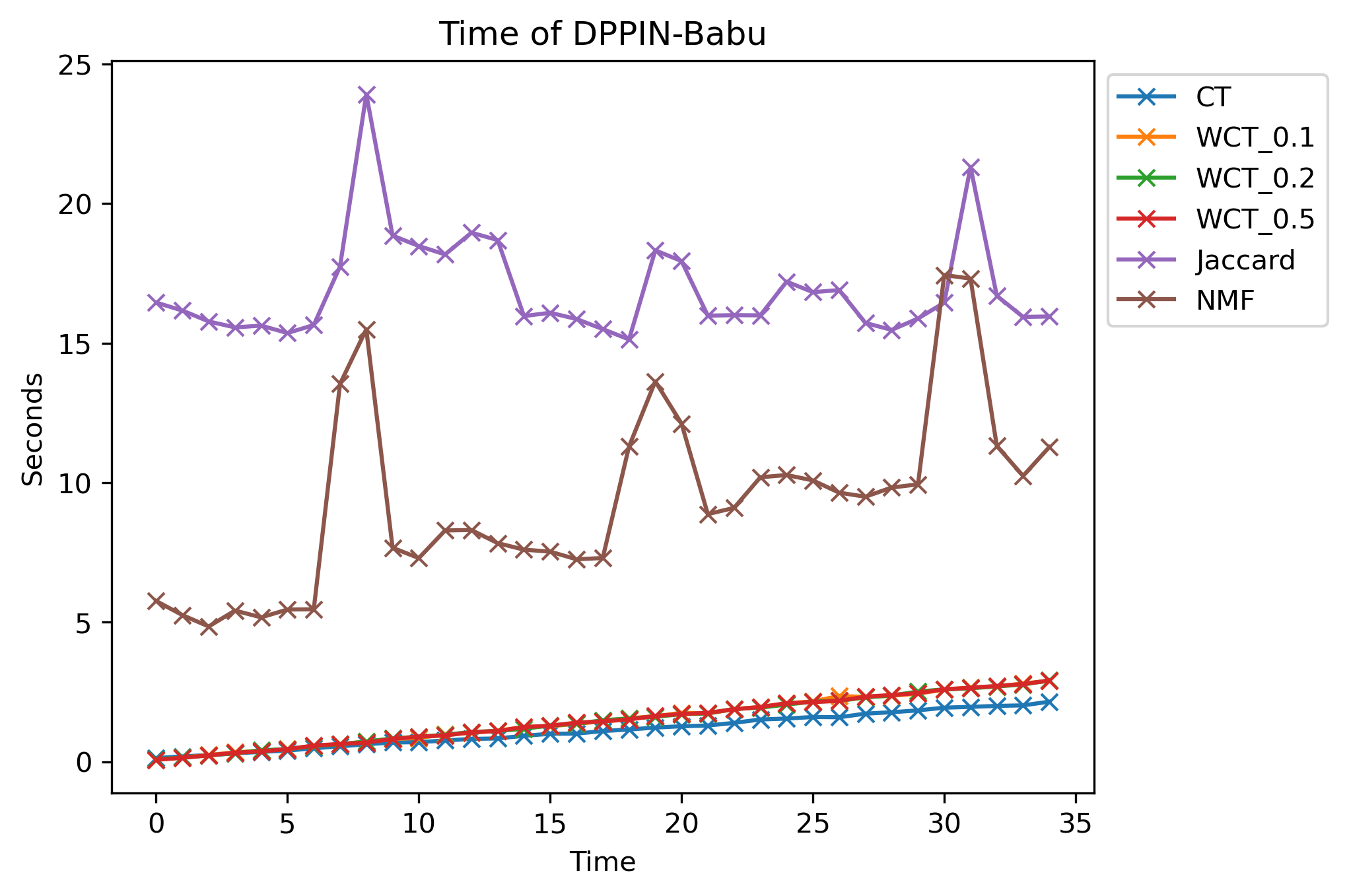}
    \caption{Time consumption in seconds.}
  \end{subfigure}
  \begin{subfigure}[b]{0.45\textwidth}
    \centering
    \includegraphics[width=\textwidth]{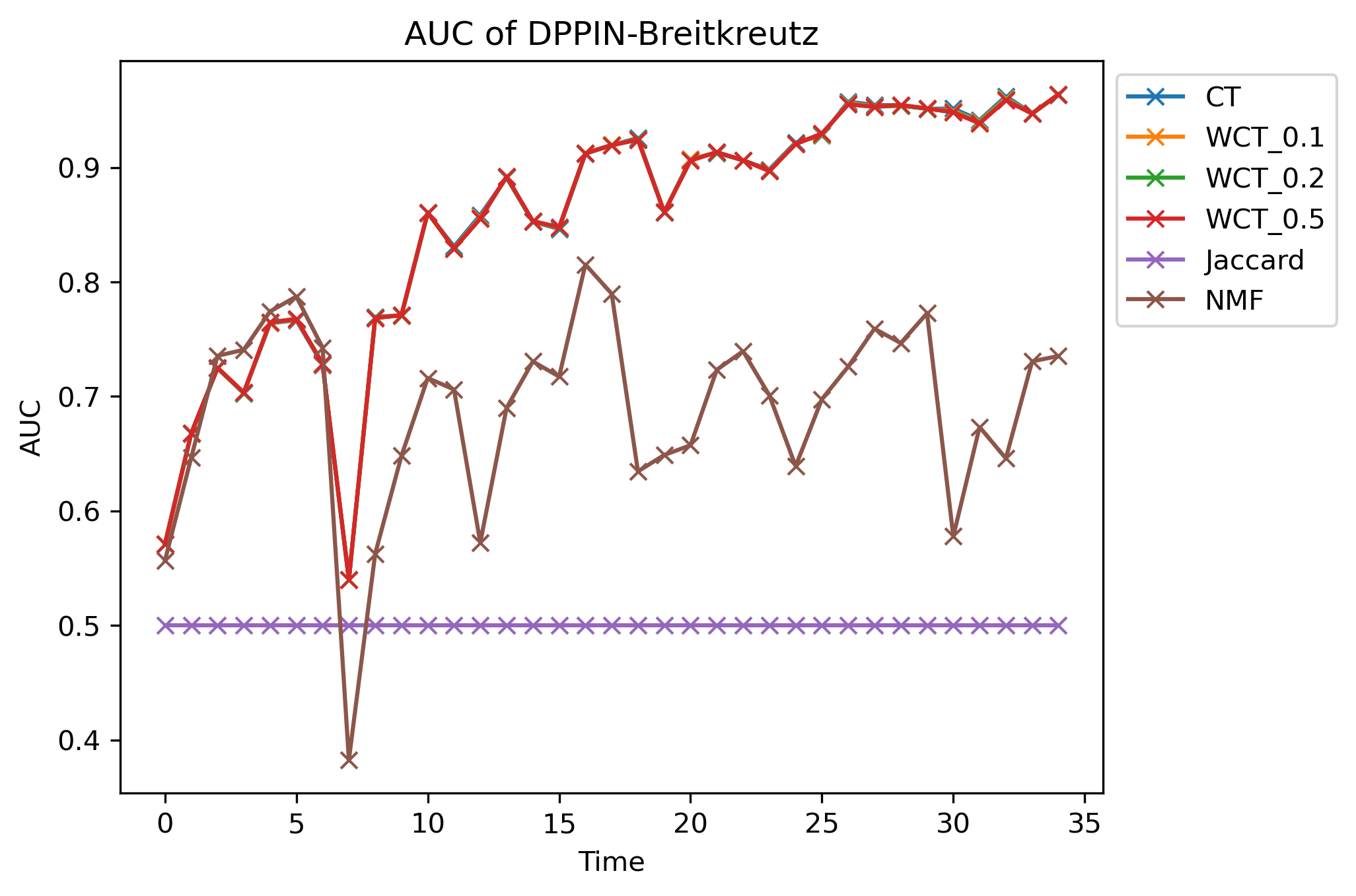}
    \caption{AUC performance of DPPIN-Breitkreutz network.}
  \end{subfigure}
  \hfill
  \begin{subfigure}[b]{0.45\textwidth}
    \centering
    \includegraphics[width=\textwidth]{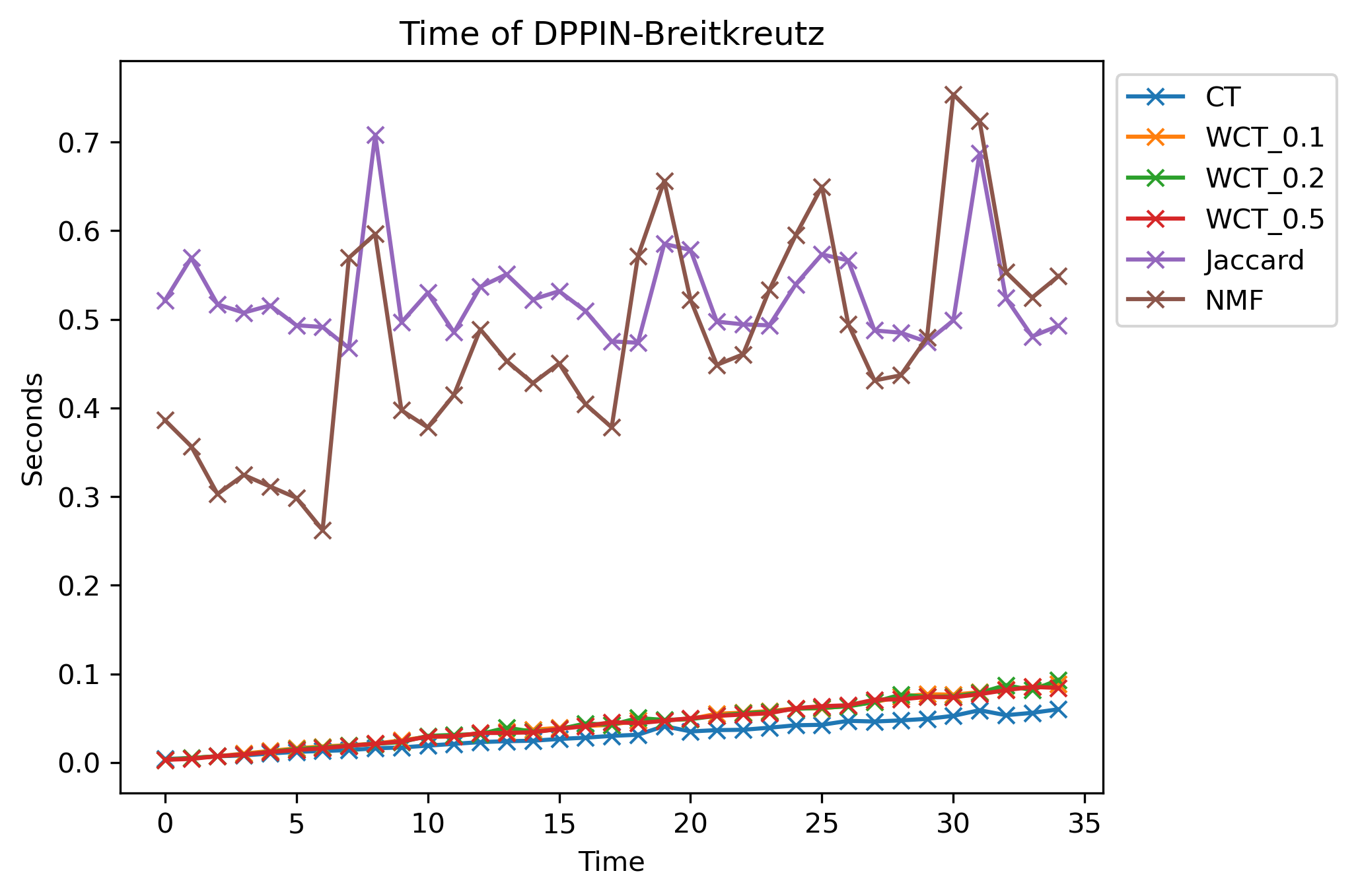}
    \caption{Time consumption in seconds.}
  \end{subfigure}
  \begin{subfigure}[b]{0.45\textwidth}
    \centering
    \includegraphics[width=\textwidth]{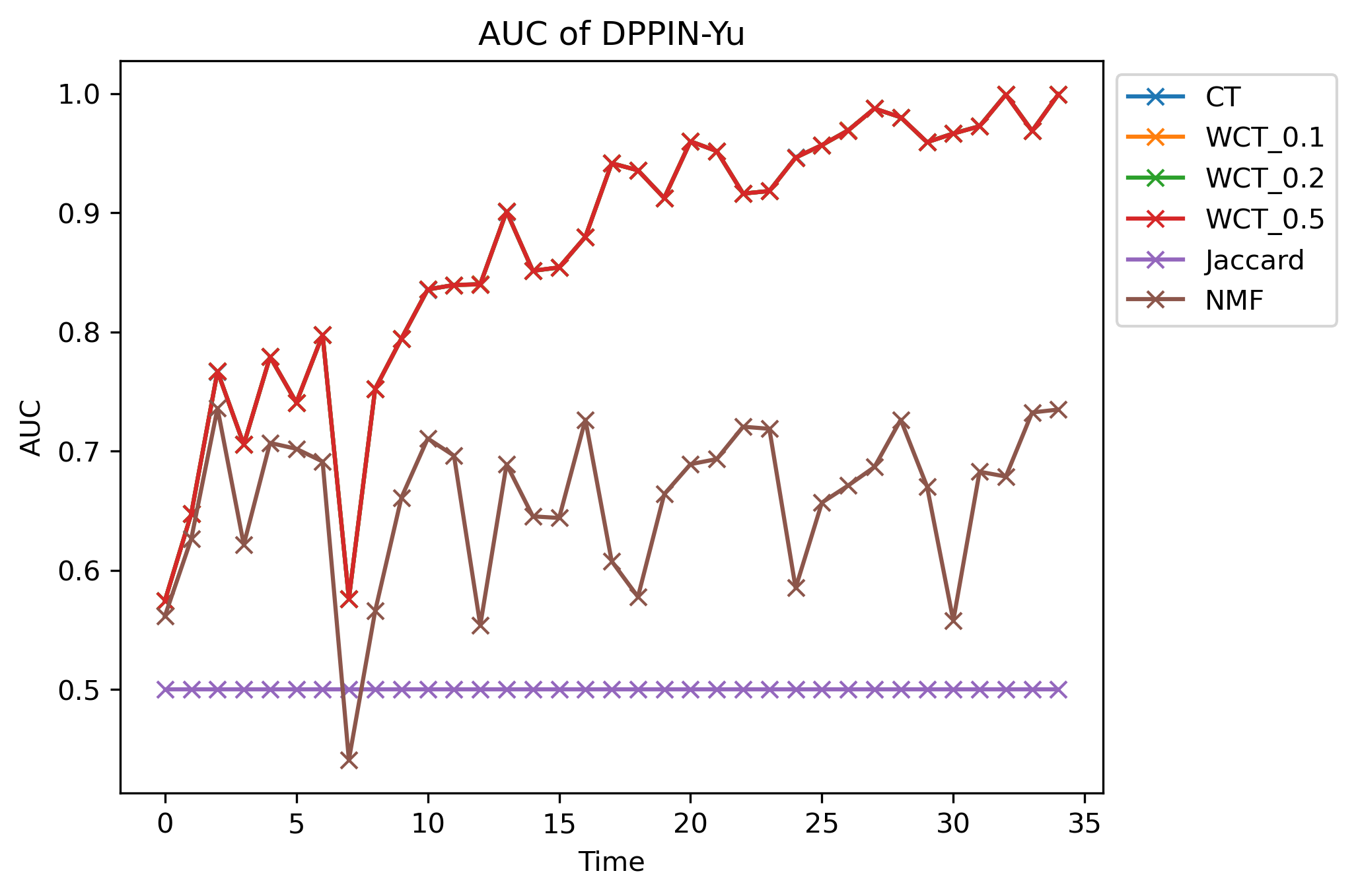}
    \caption{AUC performance of DPPIN-Yu network.}
  \end{subfigure}
  \hfill
  \begin{subfigure}[b]{0.45\textwidth}
    \centering
    \includegraphics[width=\textwidth]{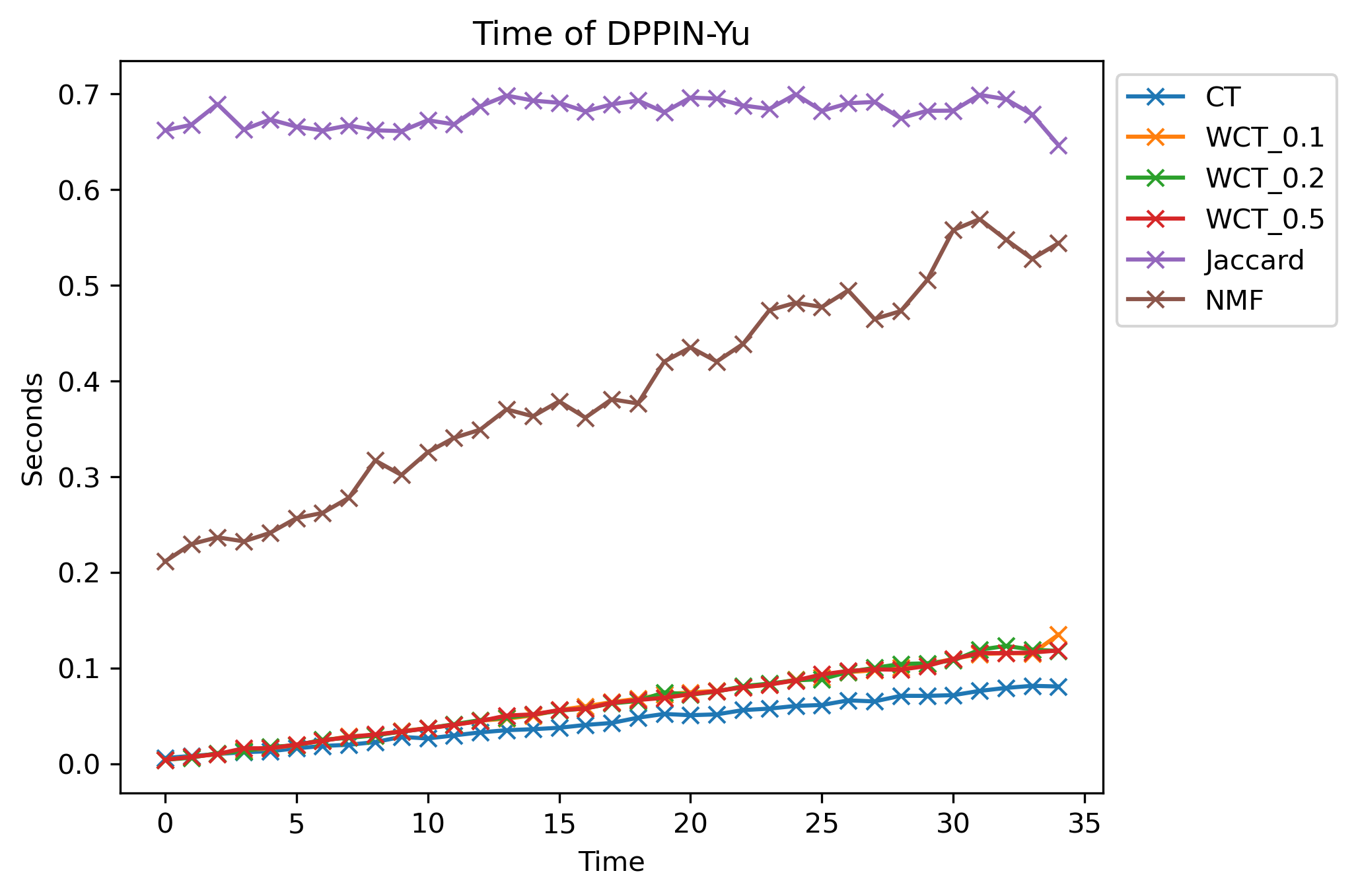}
    \caption{Time consumption in seconds.}
  \end{subfigure}
  \caption{Performance evaluation and time consumption of the four different models (CT, WCT, Jaccard, and NMF) implemented on three dynamic networks of PPIN (Babu, Breitkreutz, and Yu). Several models have very close performance. We reported three different performances of WCT with $\theta \in \{0.1,0.2,0.5\}$.}
\label{fig:temporal}
\end{figure}

\subsection{Performance Evaluation of the temporal models}


Analyzing the performance of link prediction models across the \textit{DPPIN-Babu}, \textit{DPPIN-Breitkreutz}, and \textit{DPPIN-Yu} temporal networks reveals distinct trends (see Figs. \ref{fig:temporal}a, \ref{fig:temporal}c, \ref{fig:temporal}e). The Collapsed Tensor (CT) model demonstrates a steady performance, indicating consistent predictive capability over time. Weighted Collapsed Tensor (WCT) models exhibit increased AUC with higher weights, particularly notable in the \textit{DPPIN-Babu} network, suggesting a stronger alignment with the network's temporal dynamics. In contrast, the Jaccard and Non-negative Matrix Factorization (NMF) models consistently show lower AUC scores across all networks, implying limited effectiveness in temporal link prediction. Overall, WCT models emerge as the most effective, balancing robustness and adaptability to temporal changes. We note that different weights show no impact on the metric.

The time analysis of link prediction models (see Figs. \ref{fig:temporal}b, \ref{fig:temporal}d, \ref{fig:temporal}f) on temporal networks indicates that the Collapsed Tensor (CT) method maintains consistent and low computational time across all networks, highlighting its efficiency. The Jaccard and Non-negative Matrix Factorization (NMF) methods exhibit sporadic spikes in computational time, suggesting variability in processing depending on the network's state at different time points. Overall, the CT method stands out for its temporal efficiency in link prediction tasks across the examined networks.



\begin{table}[h!]
\small
\caption{Biological and medical databases and datasets used in the literature and the experiments conducted in this research.
}

\centering
\scalebox{0.7}{

\begin{tabular}{p{2cm} p{2cm} p{3cm} p{10cm}} 
 \hline
\textbf{Publication}	&	\textbf{Database / Dataset}	&	\textbf{Link} & \textbf{Description}	\\

Knox et al. \cite{knox2010drugbank}, 2011	&	DrugBank	&	\url{https://go.drugbank.com/}	&	An extensive data on drugs and drug targets, including their chemistry, action, pharmacology, and the diseases, proteins, genes, and organisms they affect. It offers in-depth information on drug pathways, pharmacogenomics, adverse reactions, and more.	\\

Kuhn et al. \cite{kuhn2010side}, 2010	&	SIDER	&	\url{http://sideeffects.embl.de/}	&	A comprehensive, computer-readable resource that connects 888 drugs to 1450 distinct side effect terms. The SIDER dataset encompasses 62,269 drug-side effect pairs, with about 70\% of the drugs listed having between 10 and 100 different side effects. This dataset highlights the distribution of side effects across different drugs.\\

Prasad et al. \cite{keshava2009human}, 2010	&	HPRD	&	\url{http://www.hprd.org/}	&	A database about human proteome, including PPI, post-translational modifications, protein subcellular localization, and more.	\\

He, Y. et al. \cite{he2018link}, 2018	&	EEG of epilepsy patients	&	\url{https://epilepsy.uni-freiburg.de/}	&	A dataset developed from extensive EEG recordings of 8 patients with different types of pharmocoresistant focal epilepsy, capturing intricate details of seizure activities and neural firings. These recordings were processed into weighted networks using phase locking values (PLV) derived from subband EEG oscillations, providing a comprehensive framework for analyzing the initiation, progression, and termination of seizures in the brain.\\

Prieto et al. \cite{prieto2022disnet}, 2022	&	DISNET	&	\url{https://disnet.ctb.upm.es}	&	An extensive collection of biomedical data aimed at facilitating drug repurposing and disease understanding through the creation and analysis of complex networks. It integrates heterogeneous biomedical information(e.i. symptoms, signs, and other medical data) to construct customizable disease networks.	\\

Hindorff, L. et al. \cite{hindorff2009catalog}, 2012	&	GWAS catalog	&	\url{http://www.genome.gov/gwastudies}	&	A centralized database of various genome-wide association studies, aimed to identify the association between genetic variations and traits or diseases.	\\

Havugimana et al. \cite{havugimana2012census}, 2012	&	hsaPPI	&		&	Human PPI network created by merging biochemical fractionation data with spectrometric profiling and computational filtering.	\\
Giurgiu et al. \cite{giurgiu2019corum}, 2009	&	Corum	&	\url{http://mips.helmholtz-muenchen.de/corum/}	&	A collection of experimentally verified mammalian protein complexes and their interactions.	\\
Lee et al. \cite{lee2011prioritizing}, 2011	&	HumanNet	&		&	A comprehensive network that connects human genes on a genome-wide scale, combined from 21 extensive genomics and proteomics datasets, with the weight assigned indicating the strength of the evidence supporting each interaction's identification.	\\
Franceschini et al. \cite{franceschini2012string}, 2012	&	String	&	\url{http://string-db.org/}	&	The physical and functional interactions of genes/proteins from diverse sources and organisms. Edges are weighted with a probabilistic confidence score.	\\
Schmitt et al. \cite{schmitt2014funcoup}, 2014	&	FunCoup	&	\url{http://funcoup.sbc.su.se/}	&	The interaction network of genes and proteins of diverse sources using Bayesian approaches.	\\
Linghu et al. \cite{linghu2009genome}, 2009	&	FLN	&		&	A comprehensive interaction of genes (from 6 organisms) and their links association with a common biological process.	\\
Huntley et al. \cite{huntley2015goa}, 2015	&	GO	&	\url{http://www.ebi.ac.uk/GOA}	&	A comprehensive resource that provides standardized and structured information about the functions, processes, and cellular locations of genes and gene products. It shows the biological roles of genes and how they contribute to various cellular processes and functions. An example link between a pair of genes can be weighted based on the number of GO terms in common.	\\

Kuhn, M. et al. \cite{kuhn2012stitch}, 2012	&	STITCH3	&	\url{ http://stitch.embl.de/}	&	An interaction networks of proteins and chemicals extracted from a diverse set of experiments, various databases, and literature, integrating multiple sources of experimental and manually curated data. The database encompasses over 300,000 chemicals and 2.6 million proteins from 1133 different organisms, providing a broad scope of interaction information. 	\\

G{\"u}ndo{\u{g}}an, E. and Kaya, B. \cite{gundougan2017link}	&	Drugs	&	\url{https://www.drugs.com/medical_conditions.html}	&	\textit{Drugs.com} offers detailed information on a wide range of diseases and conditions. It provides comprehensive insights, including symptoms, diagnosis, and treatment options for each listed condition. 	\\

Pi{\~n}ero, J. et al. \cite{pinero2016disgenet}, 2016	&	DisGeNET	&	\url{http://www.disgenet.org/}	&	An extensive collection of gene-disease associations (GDAs). As of its version 5.0, DisGeNET contains 561,119 GDAs, linking 17,074 genes to 20,370 diseases, disorders, traits, and clinical or abnormal human phenotypes.	Moreover, DisGeNET also includes specific data collections such as the COVID-19 data set in version 5, which contains 1843 genes, 4018 diseases, 211 variants, and phenotypes, derived from over 49,410 publications. \\

Y. Li et al. \cite{li2014hmdd}, 2014	&	HMDD	&	\url{http://cmbi.bjmu.edu.cn/hmdd}	&	A comprehensive collection of experimentally supported associations between human microRNA (miRNAs) and various diseases.\\

T. Cui et al. \cite{cui2018mndr}, 2018	&	MNDR	&	\url{www.rna-society.org/mndr/}	& A network of association between \textit{long non-coding RNA}s (lncRNA) and different diseases. Edges are scored based on the quality and quantity of reference studies.	\\

Jiang, Q et al. \cite{jiang2015lncrna2target}, 2015	&	LncRNA2Target	&	\url{http://www.lncrna2target.org/}	&	This database covers both human and mouse lncRNAs. As of version 2.0, LncRNA2Target included 152,137 lncRNA–target associations, which were compiled from 1,047 papers and 224 datasets. 	\\

\end{tabular}
}
\label{tab:datasets1}
\end{table}

\begin{table}[h!]
\small
\caption{Additional biological and medical databases and datasets used in the literature and the experiments conducted in this research. 
}

\centering
\scalebox{0.7}{

\begin{tabular}{p{2cm} p{2cm} p{3cm} p{10cm}} 
 \hline
\textbf{Publication}	&	\textbf{Database / Dataset}	&	\textbf{Link} & \textbf{Description}	\\

Chen, G et al. \cite{chen2012lncrnadisease}, 2012.	&	LncRNADisease	&	\url{http://cmbi.bjmu.edu.cn/lncrnadisease}	&	An extensive compilation of associations between lncRNAs (including circular RNAs) and various diseases, enriched by a substantial addition of 25,440 new lncRNA-disease associations and curated from a wide range of literature sources. This advanced iteration enhances its utility with features like transcriptional regulatory relationships among lncRNA, mRNA, and miRNA, confidence scores for associations, and detailed curation of lncRNA interactions at multiple molecular levels, solidifying its role as a critical tool for research into the intricate roles of lncRNAs in disease.	\\

Miao, Y. et al. \cite{miao2018lncrnasnp2}, 2018	&	lncRNASNP2	&	\url{http://bioinfo.life.hust.edu.cn/lncRNASNP2}	&	An extensive information on functional \textit{single nucleotide polymorphisms} (SNPs) and mutations in human and mouse long non-coding RNAs (lncRNAs). This database offers a detailed repository of SNPs in lncRNAs, their effects on lncRNA structure, and the expression of quantitative trait loci (eQTLs) of lncRNAs, making it a valuable tool for understanding how these genetic variations affect lncRNA function and regulation. \\

Rossi, RA and Ahmed, NK,\cite{nr}, 	& NetworkRepository &	\url{https://networkrepository.com/}	& An interactive data repository with a web-based platform for visual interactive analytics. It provides an extensive collection of diverse networks. 1) \textit{ENZYMES\_g296}: This is a network of cheminformatics., 2) \textit{bn-mouse\_visual-cortex\_2}: This is a brain network. 3) \textit{bn-macaque-rhesus\_brain\_1}: A connectome (or neural connections) existed in the brain of rhesus macaque monkeys. 4) \textit{bio-yeast-protein-inter}: The network of protein-protein interaction in yeast. 5)\textit{ bio-grid-mouse}.	6) \textit{bio-celegans-dir}. \\




Kibbe, WA. et al. \cite{kibbe2015disease}, 	&Se-DoDecagon \_sidefx	&	\url{https://snap.stanford.edu}, \cite{zitnik2018modeling}	&	This dataset organizes drug side-effects into various categories, each corresponding to different classes of diseases. The side effects are essentially additional ailments that arise alongside a patient's main medical condition, for which the medication is intended. Classification of these side effects is based on their origin and the bodily systems they impact, aligning them with respective disease classes. The data regarding these disease classes is derived from the Disease Ontology.\\

       &   PDN	&	\url{https://www3.nd.edu/~dial/data/diseasenetworks/}	&	A Phenotypic Disease Network (PDN) is built using actual patient data, where diseases are represented as nodes and the edges reflect the co-occurrence or co-morbidity of these diseases.\\

Fu, D. and He, J. \cite{fu2022dppin}, 	&PPIN	&	\url{https://github.com/DongqiFu/DPPIN/tree/main}	&	A collection of twelve individual dynamic network datasets, each representing dynamic protein-protein interactions within yeast cells, but at different scales.\\

 \hline

\end{tabular}
}
\label{tab:datasets2}
\end{table}

\section{Challenges and Future Directions}
We studied the link prediction approaches applied to complex networks of myriad biological entities (enumerated in Tables. \ref{tab:datasets1} and \ref{tab:datasets2}). In this section, we enumerate the key considerations in the application of future LP models to network biology.

\subsection{Missing Information, Bias and Noise}

A major challenge in the application of link prediction on biological networks stems from the incompleteness of information. Several interactions within the existing protein-protein interaction networks and gene regulatory networks are unknown~\cite{guo2022challenges}. Many would argue that it is the very task of link prediction algorithms to infer these missing associations. However, link prediction solutions employ semi-supervised learning, where known interactions are used to train the models before reporting unknown associations~\cite{muzio2021biological}. Thus, the missing associations are likely to create bias in the networks and have an adverse effect on the subsequent predictions. The fact that the networks are not complete precludes us from deriving inferences from their known topological properties. For instance, the knowledge that signaling, transcriptional, and protein networks are scale-free~\cite{barabasi2004network,albert2005scale} does not warrant that the sampled subnetworks are scale-free as well~\cite{heath2009computational}.

In the course of this survey, we have presented link prediction approaches that leverage machine learning to learn low-dimensional vector representations based on network relationships. Once the embeddings are created, the networks lend themselves to the application of traditional supervised machine learning methods, such as Support Vector Machines and decision trees, etc.~\cite{su2020network}. The similarity between the vectors is often used as a measure of link likelihood between two biological entities. However, there are innate assumptions in the embedding models~\cite{guo2022challenges}. If, for instance, the models are based on transitivity and semantic matching, it may be difficult for them to capture symmetry and inversions. Along the same lines, the embedding strategy needs to be relevant in the context of a biological question. If nodes are embedded based on local network topology, then the prediction cannot be expected to capture global features. Biological information, such as the structure of proteins and protein complexes, the layout of signaling pathways, or the higher-order organization of organelles, etc. are likely to be lost~\cite{ideker2017network}. Moreover, the noise and sparseness in the biomedical datasets diminish the predictive capabilities of the embedding approaches. 

\subsection{Interpretability}

A significant challenge in link prediction on biomedical data is the prediction of the interpretation of the association between a pair of entities. In other words, let us assume that a drug and a gene are deemed to be associated; what does this association indicate from a biological standpoint? The drug can act as a modulator, blocker, antagonist, activator, etc. Existing embedding techniques do not often fail to provide biological context to the predicted associations\cite{guo2022challenges}. To be of utility to biologists and clinicians, the network construction phase should emphasize incorporating domain knowledge and the link prediction metrics should be able to accurately predict the nature of the association between entities~\cite{zhou2021progresses}. Once again, accurate link prediction rests on overcoming computational challenges in network structure learning. For instance, standard approaches like Bayesian inference become intractable as the number of variables increases~\cite{zhang2014network}.

\subsection{Network Motifs}\label{sec:disc_motifs}
 
In the future, biological and biomedical networks are going to become more heterogeneous, aggregating data from multiple sources with biological nuances, such as cell-type specificity, spatial and temporal resolution, or environmental factors~\cite{guo2022challenges}. The resultant networks are topologically complex, characterized by weights, link directionality, multiple layers, etc., making it difficult to assimilate information. Future prediction models will need to adapt LP approaches applied to other domain networks in the realm of biological and biomedical networks to cope with temporal dynamics, such as activation and perturbation, as well as scalability challenges in large-scale biological systems. Since local metrics relying on triangle closing or triadic closure, path counting, and graph kernels may be insufficient to tackle this complexity, the community may be likely to explore deep machine learning-based dynamic embedding techniques~\cite{koutrouli2020guide} as well as higher-order analysis that takes into account network substructures (or \textit{motifs})~\cite{roy2020motifs}. There have been a few efforts to predict links based on the motif participation of nodes~\cite{abuoda2020link} or the presence or absence of motifs~\cite{roy2023inferring}. A recent embedding algorithm combines motifs with deep learning approaches to predict associations~\cite{wang2020model}. Finally, others have combined network reconstruction to meet link prediction goals~\cite{wang2017kernel,wu2019enhancing}. 

\subsection{Heterogeneity in Knowledge Graphs}
As discussed in Sec. \ref{sec:disc_motifs}, biological systems are an amalgamation of diverse interactions including molecular interactions, gene functions, disease associations, etc. Knowledge graphs are emerging as an effective tool for researchers to access a holistic view of the biological landscape, facilitating data-driven insights, visualization, and hypothesis generation~\cite{mohamed2021biological}. The application of knowledge graphs ranges from literature mining and biological pathways analysis to personalized medicine. Future LP techniques need to overcome the following intrinsic challenges of biological and biomedical knowledge graphs to make accurate predictions: (a) \textit{Imbalanced data.} A biomedical knowledge graph may possess some common and some rare associations. This class imbalance is likely to lead to biased predictions, (b) \textit{Semantic heterogeneity.} Biological entities are often described using diverse ontologies and terminologies. Thus, link prediction models must handle the semantic heterogeneity by mapping or aligning different entity types and relationships, and (c) \textit{Incorporating multi-omics data.} Knowledge graphs include multi-omics data, such as genomics, proteomics, and metabolomics. LP models must assimilate data from these diverse sources to make biologically relevant predictions.

\section{Conclusion}
In this review, we explored the applications of link prediction (LP) methodologies within biological and biomedical networks comprising diseases, genes, proteins, RNA, microbiomes, drugs, and neurons. These LP approaches surveyed in this study fall under local, centrality, and embedding-based categories, each with its strengths and limitations. We conducted a meticulous performance evaluation of the most widely used LP methods, leveraging biological network datasets to provide insights into their efficacy and applicability. Finally, we highlighted the challenges such as noise, bias, data sparseness, and interpretability intrinsic to biological systems and how resolving them remains a crucial undertaking for future LP models in inferring biologically relevant interactions.

\section*{Acknowledgment}
This work was partially supported by 5R21MH128562-02 (PI: Roberson-Nay), 5R21AA029492-02 (PI: Roberson-Nay), CHRB-2360623 (PI: Das), NSF-2316003 (PI: Cano), VCU Quest (PI: Das) and VCU Breakthroughs (PI: Ghosh) funds awarded to P.G.

\bibliographystyle{unsrt}
\bibliography{sample-base}

\end{document}